\documentclass[%
 reprint,
 superscriptaddress,
 amsmath,amssymb,
prresearch
]{revtex4-2}

\usepackage{graphicx}
\usepackage{dcolumn}
\usepackage{bm}
\usepackage{hyperref}
\usepackage{bbm}
\usepackage{braket}
\usepackage{multirow}
\usepackage{amsmath,amssymb}
\usepackage{tikz} 
\usetikzlibrary{fit, backgrounds,shadows,3d, calc, arrows.meta, positioning, decorations.pathreplacing}
\usetikzlibrary{decorations.pathmorphing}
\usepackage{zx-calculus}
\usepackage[ruled,vlined,linesnumbered]{algorithm2e}
\usepackage{xcolor}

\definecolor{myOrange}{HTML}{F9AD81}    
\definecolor{myBlue}{HTML}{6D9EEB}      
\definecolor{myDarkBlue}{HTML}{2F5597}  
\definecolor{myGrey}{HTML}{E7E6E6}      
\definecolor{gridGrey}{HTML}{B7B7B7}    
\definecolor{ibmGrey}{HTML}{D0D0D0}
\definecolor{slowRed}{HTML}{C00000}     
\definecolor{fastGreen}{HTML}{008000}   
\definecolor{nord-polar-0}{HTML}{2E3440}  
\definecolor{nord-polar-1}{HTML}{3B4252}
\definecolor{nord-polar-2}{HTML}{434C5E}
\definecolor{nord-polar-3}{HTML}{4C566A}
\definecolor{nord-snow-0}{HTML}{D8DEE9}
\definecolor{nord-snow-1}{HTML}{E5E9F0}
\definecolor{nord-snow-2}{HTML}{ECEFF4}
\definecolor{nord-frost-0}{HTML}{8FBCBB}  
\definecolor{nord-frost-1}{HTML}{88C0D0}  
\definecolor{nord-frost-2}{HTML}{81A1C1}  
\definecolor{nord-frost-3}{HTML}{5E81AC}  

\definecolor{nord-aurora-red}{HTML}{BF616A}
\definecolor{nord-aurora-orange}{HTML}{D08770}
\definecolor{nord-aurora-yellow}{HTML}{EBCB8B}
\definecolor{nord-aurora-green}{HTML}{A3BE8C}
\definecolor{nord-aurora-purple}{HTML}{B48EAD}

\newcommand{\cnn}{\textsc{CNN}}
\newcommand{\LER}{\mathrm{LER}}
\newcommand{\ratio}[2]{\rho_{\mathrm{#1/#2}}}
\newcommand{\gain}[2]{\mathcal{G}_{\mathrm{#1}\leftarrow\mathrm{#2}}}

\usepackage{tabularx, booktabs}
\begin{document}

\title{Calibration-Conditioned FiLM Decoders for Low-Latency Decoding of Quantum Error Correction Evaluated on IBM Repetition-Code Experiments}

\author{Samuel Stein}
\email{samuel.stein@pnnl.gov}
\affiliation{%
  Physical and Computational Sciences, Pacific Northwest National Laboratory, Richland, Washington, USA
}

\author{Shuwen Kan}
\affiliation{%
  Physical and Computational Sciences, Pacific Northwest National Laboratory, Richland, Washington, USA
}
\affiliation{Department of Computer and Information Sciences, Fordham University, USA}

\author{Chenxu Liu}
\affiliation{%
  Physical and Computational Sciences, Pacific Northwest National Laboratory, Richland, Washington, USA
}
\author{Adrian Harkness}
\affiliation{%
  Physical and Computational Sciences, Pacific Northwest National Laboratory, Richland, Washington, USA
}
\affiliation{Department of Industrial and Systems Engineering, Lehigh University}

\author{Sean Garner}
\affiliation{%
  Physical and Computational Sciences, Pacific Northwest National Laboratory, Richland, Washington, USA
}
\affiliation{Department of Electrical \& Computer Engineering, University of Washington, USA}

\author{Zefan Du}
\affiliation{Department of Computer and Information Sciences, Fordham University, USA}

\author{Yufei Ding}
\affiliation{University of California San Diego, San Diego, USA}

\author{Ying Mao}
\affiliation{Department of Computer and Information Sciences, Fordham University, USA}

\author{Ang Li}
\affiliation{%
  Physical and Computational Sciences, Pacific Northwest National Laboratory, Richland, Washington, USA
}
\affiliation{Department of Electrical \& Computer
Engineering, University of Washington, USA}

\begin{abstract}
Real-time decoding of quantum error correction (QEC) is essential for enabling fault-tolerant quantum computation. A practical decoder must operate with high accuracy at low latency, whilst remaining robust to spatial and temporal variations in hardware noise. We introduce a hardware-conditioned neural decoder framework designed to exploit the natural separation of timescales in superconducting processors, where calibration drifts occur over hours while error correction requires microsecond-scale responses. By processing calibration data through a graph-based encoder and conditioning a lightweight convolutional backbone via Feature-wise Linear Modulation (FiLM), we decouple the heavy processing of device statistics from the low-latency syndrome decoding.

We evaluate this approach using the \textbf{1D repetition code} as a testbed on IBM \textit{Fez}, \textit{Kingston}, and \textit{Pittsburgh} processors, collecting over 2.7 million experimental shots spanning distances up to $d=11$. We demonstrate that a single trained model generalizes to unseen qubit chains and new calibration data acquired days later without retraining. On these unseen experiments, the FiLM-conditioned decoder achieves up to a $11.1\times$ reduction in logical error rate relative to modified minimum-weight perfect matching. We observe that by employing a network architecture that exploits the highly asynchronous nature of system calibration and decoding, hardware-conditioned neural decoding demonstrated promising, adaptive performance with negligible latency overhead relative to unconditioned baselines.
\end{abstract}

\maketitle


\section{Introduction}\label{sec:intro}

Quantum computers are inherently noisy, and this noise limits their ability to perform meaningful large-scale computation. Quantum error correction (QEC) is the leading path to suppressing physical errors to the levels required to execute computationally meaningful quantum algorithms \cite{lidar2013quantum,beverland2022assessing}. By encoding logical qubits into a larger Hilbert space and repeatedly performing syndrome extraction, fault-tolerant protocols can reduce logical error rates quadratically with increasing code distance~\cite{terhal2015quantum,gottesman1997stabilizer}. Realising this promise hinges not only on hardware but also on the accompanying classical stack. A decoder must correctly predict the required corrections at the cadence of repeated syndrome extraction such that we can drive computation \cite{fowler2012towards}. If the decoder lags or produces incorrect corrections, the code-decoder setting forfeits its utility.

Within the hardware landscape of quantum computing, superconducting processors constitute one of the leading paths to fault tolerance. However, they face operational challenges pertinent to decoding: qubit performance is spatially heterogeneous, noise channels are highly asymmetric, and the intrinsic speed of superconducting processors necessitates decoding latencies on the microsecond timescale. To maintain fault tolerance under these conditions, decoders must be high-throughput, accurate, and capable of adapting to heterogeneous noise profiles without the operational cost of constant retraining. Accordingly, it is important to assess decoders in experimental settings, since commonly used simulated noise models often omit precisely the latent hardware dynamics that challenge decoder performance.

Publicly available IBM superconducting devices currently feature a heavy-hex lattice topology. Connectivity is restricted to degree-2 and degree-3, which precludes the native embedding of high-weight stabilizer codes, such as the surface code, without significant compilation overhead. Consequently, the repetition code serves as a strong candidate for studying decoding dynamics on native hardware~\cite{chiaverini_realization_2004,dicarlo_preparation_2010,schindler_experimental_2011,wootton2018repetition,besedin2025realizing}. Although the repetition code protects only a single Pauli basis, it shares the stabilizer formalism foundational to all stabilizer codes, including CSS codes~\cite{calderbank1996good}. Developing decoders that satisfy hardware operational requirements while performing well on these codes is a prerequisite for scalable QEC.

A practical decoder faces three simultaneous demands: (i) decoding throughput to match computational demand of the underlying code, (ii) accuracy in matching error syndromes to the underlying error, and (iii) adaptability to spatiotemporally varying noise and hardware without bespoke retuning and redesign. Classical analytic decoders such as minimum-weight perfect matching (MWPM) set strong baselines for topological codes~\cite{fowler2012proof}, but they rely on an approximated error model, explicitly derived from hardware statistics; deviations from idealized assumptions such as bias, inhomogeneity and correlated errors all can degrade performance.

Motivated by these trade-offs, multiple recent experiments have turned to machine-learning decoders that infer noise structure directly from experiment data \cite{bausch2024learning,google2025quantum}. Convolutional models exploit locality and translational invariance of detector events, while more expressive transformer \cite{wang2023transformer} and graph models \cite{lange2025data} can capture longer-range space–time correlations. For instance, a recurrent transformer-based decoder trained and then fine-tuned on real syndromes outperformed state-of-the-art algorithmic decoders on Google’s Sycamore data at distances $d=3$ and $d=5$, attaining a $2.748\%$ logical error rate (LER) against the MWPM-Correlated $3.597\%$ at $d=5$ ~\cite{bausch2024learning}. Graph-neural decoders trained on detector graphs have surpassed matching under circuit-level noise in simulation and reached parity with MWPM on experimental repetition-code data~\cite{lange2025data}. These results are compelling, though many learned decoders are trained for a single fixed device/noise distribution and require retraining or fine-tuning, and maintaining low latency can restrict the model's complexity and architecture. For a decoder to scale, it must generalize beyond the specific experiments on which it is trained. 

We pursue a route to decoder adaptability that leverages the separation of timescales between hardware drift and decoding. Device characteristics such as coherence times and gate fidelities drift over hours or days \cite{carroll2022dynamics}, whereas syndrome decoding in superconducting systems operates at the microsecond scale. To exploit this, we employ \emph{Feature-wise Linear Modulation} (FiLM)~\cite{perez2018film}. FiLM applies channel-wise transformations to intermediate features based on external side information. By using the varying calibration data to generate these modulation parameters asynchronously, we can condition a lightweight convolutional backbone on the current device state without treating the hardware state as a real-time input. This strategy allows the decoder to adapt to spatial and temporal noise variations, while preserving the low latency lightweight neural model required for the QEC cycle.

In this work, we propose a calibration-conditioned neural decoder that couples a graph-neural system encoder with a FiLM-modulated convolutional backbone operating on detection-event tiles. The graph encoder ingests per-qubit and per-edge calibration features (e.g., $T_1/T_2$, readout assignment errors, gate error rates) and produces a latent conditioning vector that modulates the convolutional feature maps through FiLM layers. We train one model per $(d,r)$ configuration (distance $d$ and number of rounds $r$), with experiment data generated from three IBM devices and multiple contiguous physical qubit subsets, and evaluate the trained model a week later on \emph{new contiguous sets of physical qubits with new calibration data}. The model has not seen this data during training. \emph{All results reported, including the validation set, are excluded from the training process and solely used for evaluation.}

Our work draws on a large experimental corpus, comprising $2,760,704$ repetition-code shots. We utilize 400 calibration snapshots from three IBM processors, spanning code distances up to $d=11$. We evaluate the model on new contiguous sets of physical qubits with calibration data acquired one week after training. The model maintains high performance without retraining. Specifically, we observe a performance crossover at $(d,r) \approx (7,5)$ for the $Z$-basis, beyond which the FiLM-conditioned decoder consistently outperforms comparative baselines. At the largest code size $(d=11, r=11)$, the decoder achieves a $11.1\times$ improvement in logical error rate relative to hardware-informed MWPM.

In this work, we put forward three key contributions. \textit{First}, we propose a decoding framework that effectively decouples the processing of slow-varying noise characteristics from fast-path syndrome decoding. By integrating a graph-neural hardware encoder with a FiLM-modulated convolutional backbone, we enable hardware-aware inference that adapts to changing device physics at latencies comparable with unmodified convolutional neural networks. \textit{Second}, we demonstrate that this calibration-conditioned approach sustains logical-error suppression across both spatial and temporal hardware variations. In experiment, the proposed decoder generalizes to unseen qubit chains and new calibration snapshots acquired a week later, achieving up to a $7.41\times$ reduction in logical error rate relative to network without FiLM, and minimum weight perfect matching decoders. \textit{Third}, we provide our experiment dataset and raw IBM experiment archives, comprising  $\approx2.7\times10^{6}$ repetition-code shots across $400$ hardware snapshots for future works to utilize.

The remainder of this paper is organized as follows. Section~\ref{sec:background} reviews the repetition code, superconducting-hardware noise biases, and prior decoding strategies. Section~\ref{sec:architecture} details the FiLM-conditioned architecture and its integration with graph-based system features. Section~\ref{sec:results} describes the experimental pipeline and dataset, and presents performance comparisons and latency implications, including sensitivity analyses of FiLM parameters and cross-calibration transfer. Section~\ref{sec:discussion} discusses implications for scalable QEC and extensions to higher-dimensional codes, and Section~\ref{sec:conclusion} discusses natural extensions to this work and concludes.

\section{Background and problem setting}\label{sec:background}

In this section we introduce the required preliminaries and notation of our problem setting. We refer the reader to Nielsen and Chuang \cite{nielsen2010quantum}, and Gottesman \cite{gottesman1997stabilizer} for a detailed introduction to quantum error correction. 

\subsection{Quantum error correction.}
A stabilizer code encodes $k$ logical qubits into $n$ physical qubits by defining an Abelian subgroup 
$\mathcal{S}\subset\mathcal{P}_n$ of the $n$-qubit Pauli group $\mathcal{P}_n$ that excludes $-I$. 
The simultaneous $+1$ eigenspace of all elements in $\mathcal{S}$ forms the protected codespace. 
Logical operators act nontrivially on this codespace and are drawn from the normalizer 
$\mathcal{N}(\mathcal{S})=\{P\in\mathcal{P}_n\,|\, P\mathcal{S}P^{\dagger}=\mathcal{S}\}$ but not from $\mathcal{S}$ itself, 
so that $\mathcal{N}(\mathcal{S})/\mathcal{S}$ defines the logical operators. The \emph{distance} $d$ of a stabilizer code is the minimum weight of any nontrivial logical operator, that is, the smallest number of physical qubits on which a logical operator acts nontrivially. These properties are often summarized as [[n,k,d]]. A code of distance $d$ can detect up to $d-1$ and correct up to $\lfloor (d-1)/2 \rfloor$ arbitrary single-qubit errors on distinct qubits. Syndrome measurements project the system onto stabilizer eigenspaces, revealing parity information from which a decoder infers the most likely recovery operation.

\begin{figure*}[t]
    \centering
    \includegraphics[width=\textwidth]{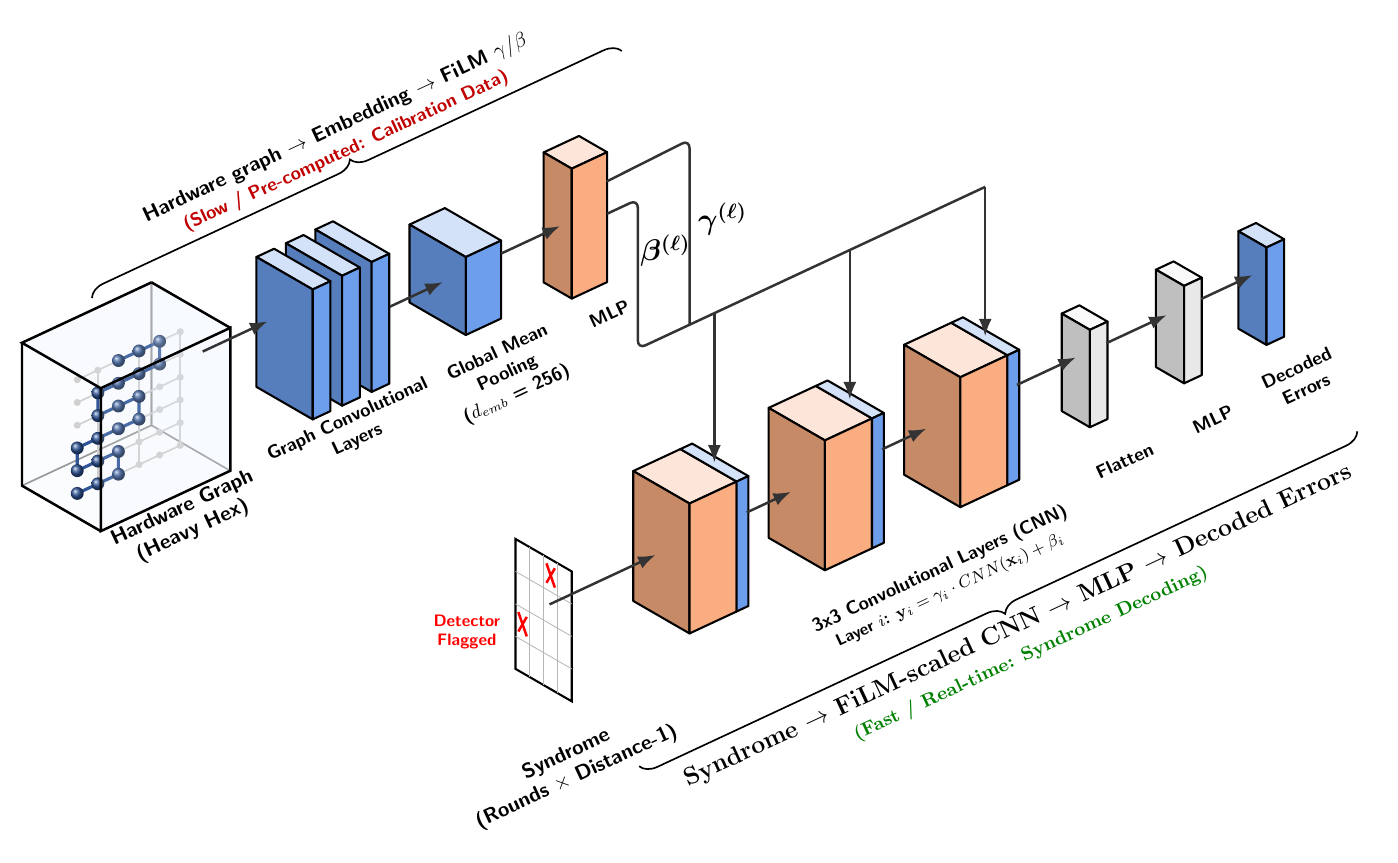}
    \caption{\label{fig:film_architecture}
    \textbf{Architecture of the Calibration-Conditioned FiLM Neural Decoder.} The framework consists of three core components. 
    \textbf{(Upper Left) Hardware Encoder:} An experimental shot defines a calibration subgraph $G=(V,E)$, extracted from the target \textbf{IBM Heavy Hex} topology device, where node features include normalized $T_1, T_2$, and gate errors. This graph is processed by a 3-layer Graph Convolutional Network (GCN) and pooled to form a latent calibration embedding $\mathbf{z} \in \mathbb{R}^{256}$.
    \textbf{(Upper Right) FiLM Generator:} A Multi-Layer Perceptron (MLP) projects the latent embedding $\mathbf{z}$ into layer-wise modulation parameters $\boldsymbol{\gamma}^{(\ell)}$ (scale) and $\boldsymbol{\beta}^{(\ell)}$ (shift).
    \textbf{(Right) Decoder Arm:} The error syndrome tensor $\chi \in \{0, 1\}^{r \times (d-1)}$ is processed by a Convolutional Neural Network (CNN). The feature maps of the CNN layers are modulated via the Feature-wise Linear Modulation (FiLM) operation: $Z^{(\ell)} = \sigma(\boldsymbol{\gamma}^{(\ell)} \odot \text{Conv}(Z^{(\ell-1)}) + \boldsymbol{\beta}^{(\ell)})$. The output CNN features are flattened and mapped to per-qubit correction probabilities via a dense output head.}
\end{figure*}

\textbf{Repetition-code} The distance-$d$ repetition code encodes one logical qubit ($k=1$) into $n$ physical data qubits. Let $X_i$ and $Z_i$ denote Pauli operators on data qubit $i$. For the bit-flip-protecting ($Z$-type) realization, the stabilizer group is generated by
\begin{equation}
\mathcal{S}_Z=\langle\, Z_i Z_{i+1}\,:\,1\le i\le d-1\,\rangle,
\end{equation}
with logical operator $\bar X=X^{\otimes d}$; the dual ($X_i\leftrightarrow Z_i$), phase-flip-protecting ($X$-type) realization interchanges $X$ and $Z$ with $\mathcal{S}_X=\langle X_i X_{i+1}\,:\,1\le i\le d-1\,\rangle$ and $\bar Z=Z^{\otimes d}$. In either case the code corrects up to $\lfloor (d-1)/2\rfloor$ errors of the protected Pauli type \cite{gottesman1997stabilizer}. Syndrome extraction is performed by measuring each stabilizer generator repeatedly using ancilla qubits. In this work, we implement this on hardware as a single dynamic circuit, where ancilla qubits are measured and reset in real-time for $r$ consecutive rounds. These measurements project the system into stabilizer eigenspaces and reveal where local parity constraints have been violated, forming the basis for the detection events that are used to decode and recover the corrected Pauli frame. 

\textbf{Syndrome extraction and detection events.}
Stabilizers are measured repeatedly for $r$ rounds using parity checks over ancilla qubits. Let $s_{t,i}\in\{0,1\}$ denote the parity outcome of stabilizer $i$ at round $t$ ($1\le t\le r$, $1\le i\le d-1$). Consecutive rounds are XORed to form detection events,
\begin{equation}
\chi_{t,i}=s_{t,i}\oplus s_{t-1,i},\qquad s_{0,i}=0,
\end{equation}
This localizes changes caused by data or measurement faults. Single faults create short, local space–time chains of detection events; longer chains indicate multiple faults or measurement errors across rounds. 

\textbf{Decoding objective.}
Decoding is tasked with inferring what errors occurred that generated the syndrome map observed \cite{ravi2023better,devitt2013quantum, hsieh2011np}. Given the full detection-event history $\chi$ and a hardware prior over Pauli errors $P(E)$, the information-theoretic optimum is the recovery map $R^\star$ that maximizes the probability of returning to the codespace without flipping the logical bit:
\begin{equation}
R^\star(\mathcal{\chi})=\arg\max_{L\in\{\mathbb{I},\,\bar X\,(\text{or }\bar Z)\}}\;
\sum_{E\in\mathcal{P}_n:EL\in\mathcal{S}} P(E\mid \chi)\,.
\end{equation}
In this work, we target a per-qubit marginal estimate. The decoder outputs $p\in[0,1]^d$ with $p_i\approx\Pr(E_i=1\mid\mathcal{\chi})$ for the protected error channel. With a threshold $\tau=0.5$, $\hat c=\mathbf{1}[p>\tau]$ updates a tracked Pauli frame, and the reported logical is extracted from the frame-adjusted final measurements. 

\subsection{IBM Hardware}

All experiments and results in this work are conducted on IBM superconducting quantum processors based on transmon qubits arranged in heavy-hex lattice topologies, specifically the \textsc{IBM Fez}, \textsc{Kingston}, and \textsc{Pittsburgh}. This architecture provides nearest-neighbor connectivity of weight 2- and 3- over a heavy-hex grid. Notably, this sparse connectivity prevents the native tiling of surface-code plaquettes or higher-weight connectivity, motivating the use of linear repetition codes as an effective low-overhead topological choice without incurring compilation or swapping overhead. IBM supports routine access to calibrated qubit parameters through IBM's open calibration interface. Each calibration snapshot includes coherence times ($T_1$, $T_2$), single- and two-qubit gate error rates, and readout assignment errors.  

\subsection{Comparable Decoders}

Neural network-based decoding has garnered attention for its potential to address the complex hardware noise profiles beyond standard algorithmic approaches. Krastanov and Jiang~\cite{krastanov2017deep} utilized deep neural networks to estimate conditional error probabilities for stabilizer codes under simulated depolarizing noise. Maskara et al.~\cite{maskara2019advantages} applied feedforward networks to topological color codes, while Varsamopoulos et al.~\cite{varsamopoulos2019comparing} investigated Long Short-Term Memory (LSTM) networks to compare high- and low-level decoding strategies in simulated environments. To address temporal correlations, Baireuther et al.~\cite{baireuther2018machine} employed recurrent neural networks trained on density-matrix simulations of surface code circuits. More recently, Lange et al.~\cite{lange2025data} utilized Graph Neural Networks (GNNs) to capture the topological dependencies of quantum codes. While these studies demonstrate the viability of machine learning for decoding, they primarily rely on simulated data. It is increasingly recognized that a key advantage of machine learning is its ability to infer noise and error dynamics directly from experimental data \cite{strikis2021learning,liao2024machine,liao2025noise}, capturing device-specific and hardware-dependent effects that idealized models often omit by training on real noisy outputs rather than on preconceived noise models  

To evaluate our calibration-conditioned approach, we compare against two baselines, a modified minimum weight perfect matching (MWPM) decoder, and a convolutional neural network (CNN) baseline. 

\paragraph{Modified MWPM (calibration-informed)}
We derive a circuit-level noise model from the transpiled schedule executed on the chosen contiguous chain and the calibration snapshot. Specifically:
\begin{itemize}
    \item \textbf{Gate errors:} each one-qubit (two-qubit) gate is followed by a Pauli-twirled depolarizing channel with probability $p_{1q}$ ($p_{2q}$) taken from the device calibration for that gate.
    \item \textbf{Measurement:} readout assignment error is inserted using the calibrated assignment probability for the measured qubit(s).
    \item \textbf{Idling:} at the end of each round, every data qubit experiences a Pauli-twirled channel whose parameters are computed from $(T_1,T_2)$ and the effective idle/reset durations in that round.
\end{itemize}

From this model we construct a detector graph whose edges connect pairs of detectors (or a detector to a boundary). Each edge is assigned a weight given by the negative log-likelihood derived from the corresponding fault probability. Running minimum-weight perfect matching on this graph returns a most-likely set of errors \cite{higgott2022pymatching}. We implement MWPM using PyMatching with edge fault probabilities derived from the circuit-level noise detector graph above. A separate detector graph and edge-weight assignment is constructed for each (device, basis, d, r, calibration snapshot) so that MWPM has access to the same calibration information as the FiLM decoder, namely T1, T2, gate error rates, and readout assignment errors.

\paragraph{CNN baseline (unconditioned).}
The CNN baseline is architecturally identical to the FiLM decoder's convolutional backbone in Figure \ref{fig:film_architecture} and output head but excludes the hardware encoder and FiLM generator. The detection tensor $\chi \in \{0,1\}^{r\times(d-1)}$ is fed directly through the convolutional blocks to produce per-qubit flip probabilities. We train one model per $(d,r)$ and basis using the same training/validation split, loss, optimizer, and inference threshold as the FiLM model. This baseline therefore controls for network architecture and training protocol, isolating the contribution of calibration conditioning. 

\section{Calibration-Conditioned FiLM Decoder Architecture}\label{sec:architecture}

\begin{algorithm}[htbp]
\caption{Calibration-Conditioned FiLM Convolutional Neural Decoder Architecture}
\label{alg:film_decoder}
\SetKwInOut{Input}{Input}
\SetKwInOut{Output}{Output}
\SetKwComment{Comment}{// }{}

\Input{
    Syndrome tensor $\chi \in \{0,1\}^{r \times (d-1)}$. \\
    Hardware graph $G$ with node calibration features.}
\Output{
    Per-qubit correction logits $\hat{y} \in \mathbb{R}^{d}$
}
\BlankLine

\tcc{1. System Encoder (GNN) - Embeds hardware properties}
$H^{(0)} \leftarrow G$ \Comment*[r]{Initial features from the graph}
\For{$\ell = 1$ \KwTo $3$}{
    $H^{(\ell)} \leftarrow \text{ReLU}(\text{GCNConv}_{\ell}(H^{(\ell-1)}))$
}
$\mathbf{z} \leftarrow \text{GlobalMeanPool}(H^{(3)})$ \Comment*[r]{Latent hardware vector $\mathbf{z} \in \mathbb{R}^{256}$}

\BlankLine
\tcc{2. FiLM Generator (MLP) - Generates conditioning parameters}
\Comment{This branch is pre-computable}
$\boldsymbol{\theta} \leftarrow W_2 \cdot \text{ReLU}(W_1 \mathbf{z} + b_1) + b_2$ \\
$(\gamma_1, \beta_1), (\gamma_2, \beta_2), (\gamma_3, \beta_3) \leftarrow \text{Partition}(\boldsymbol{\theta})$ \Comment*[r]{For CNN blocks 1, 2, 3}

\BlankLine
\tcc{3. Decoder Arm (FiLMed CNN) - Processes detection events}
$x \leftarrow \chi$ \\
\For{$\ell = 1$ \KwTo $3$}{
    \Comment{Modulate features with hardware-specific params}
    $x \leftarrow \text{ReLU}\left(\gamma_\ell \odot \text{Conv}_\ell(x) + \beta_\ell\right)$
}

\BlankLine
\tcc{4. Output Head - Generates final logits}
$x_{\text{flat}} \leftarrow \text{Flatten}(x)$ \\
$\text{logits}_{\text{flat}} \leftarrow W_{\text{head}} x_{\text{flat}} + b_{\text{head}}$ \\
$\hat{y} \leftarrow \text{logits}_{\text{flat}}$
\BlankLine
\KwRet{$\hat{y}$}
\end{algorithm}

We employ a calibration-conditioned FiLM decoder architecture motivated by three key requirements:

\begin{itemize}
    \item \textbf{Generalization:} The decoder must be robust to spatial and temporal variations in noise without requiring retraining for every specific qubit chain or calibration cycle.
    \item \textbf{Latency:} The conditioning mechanism must add minimal computational overhead at inference time.
    \item \textbf{Correlations:} The model should leverage learned noise structures from data, capturing latent dynamics that uncorrelated hardware calibration statistics may omit.
\end{itemize}

Our neural architecture is outlined in Figure~\ref{fig:film_architecture}. Each experimental shot provides a detection-event tensor and a hardware graph of the underlying qubits on which the experiment is run. The neural decoder factors into three core components (i) a hardware encoder that embeds the calibration subgraph into a learned latent vector representation, (ii) a FiLM generator that maps this latent vector context to per-convolutional layer feature map scale/bias, and (iii) a central convolutional neural pipeline that consumes detection events generated by the underlying experiment modulated by FiLM. 

For a repetition-code experiment of distance $d$, comprising $2d-1$ contiguous physical qubits and $r$ rounds, we form a detection tensor $\chi\in\{0,1\}^{ r \times (d-1)}$ (time, stabilizer index) and a calibration subgraph $G=(V,E)$. This subgraph is extracted from the backend's heavy-hex lattice, visualized as the highlighted chain in Fig~\ref{fig:film_architecture}-\textsc{Hardware Graph}, in representing the specific contiguous qubit chain on which the code is executed. Node features include normalised $(T_1,T_2)$, gate errors and readout-assignment error. The network outputs per-qubit flip probabilities $p\in[0,1]^d$ for the protected error channel. Thresholding at $\tau=0.5$ yields a binary correction frame $\hat c=\mathbf{1}[p>\tau]$ that is XORed into the tracked Pauli frame before readout. Final logical state is decoded by XORing the data qubit measurement at the end of the experiment with the tracked Pauli frame and majority voting.

\subsection{FiLM Generator \& System Encoder}

We embed the calibration subgraph $G=(V,E)$, whose vertices correspond to the
qubits in the contiguous chain and whose node features include normalized
$(T_1,T_2)$, readout assignment error, and single-qubit gate error rates, with
edge features given by two-qubit gate error rates. A three-layer graph
convolutional network (GCN) with ReLU activations and global mean pooling
produces a hardware-dependent latent embedding
\begin{equation}
    h = \frac{1}{|V|}\sum_{v\in V} 
    \mathrm{GCN}(X_V,E)_v
    \in \mathbb{R}^{d_{\mathrm{emb}}},
\end{equation}
using an embedding dimension of $d_{\mathrm{emb}}=256$.

While one could concatenate calibration features to the syndrome input, we employ FiLM as a more effective strategy to modulate internal feature maps of the underlying CNN. This allows the network to re-weight the decoder's filters based on the hardware state, adjusting the processing logic of the detection events rather than appending additional inputs, increasing per-shot compute.

The latent vector is passed through a two-layer MLP that produces FiLM
conditioning parameters for each convolutional block $\ell\in\{1,2,3\}$:
\begin{equation}
    \{(\gamma^{(\ell)},\beta^{(\ell)})\}_{\ell=1}^3
    = f_{\theta}(h), 
    \qquad 
    \gamma^{(\ell)},\beta^{(\ell)}\in\mathbb{R}^{C_\ell},
\end{equation}
where $C_\ell$ is the channel width of block $\ell$ and $(\gamma,\beta)$ broadcast over spatial dimensions. As these parameters depend only on the hardware graph $G$ and not the detection tensor $\chi$, they are precomputable.

Given FiLM parameters, the detection-event tensor $\chi\in\{0,1\}^{r\times(d-1)}$ is processed by a sequence of FiLM-modulated convolutional blocks. Each block applies the same generic transformation
\begin{equation}
    Z^{(\ell)} = 
    \sigma\!\left(
        \gamma^{(\ell)} \odot 
        \mathrm{Conv}^{(\ell)}(Z^{(\ell-1)}) 
        + \beta^{(\ell)}
    \right),
    \qquad 
    Z^{(0)} = \chi.
\end{equation}

The learned parameters $\gamma$ and $\beta$ apply a channel-wise transformation to the feature maps, amplifying or suppressing specific features based on the hardware calibration state.

After flattening the final feature map $Z^{(3)}$, a linear output head produces per-qubit flip probabilities
\begin{equation}
    \hat p_i = 
    \Pr(\text{flip on data qubit } i \mid \chi,G),
    \qquad 1\le i\le d.
\end{equation}
Thresholding at $\tau=0.5$ yields the correction frame  $\hat c=\mathbf{1}[\hat p>\tau]$, which is XORed into the tracked Pauli frame before extracting the logical state.

\begin{figure*}[htbp]
\centering
\includegraphics[width=\textwidth]{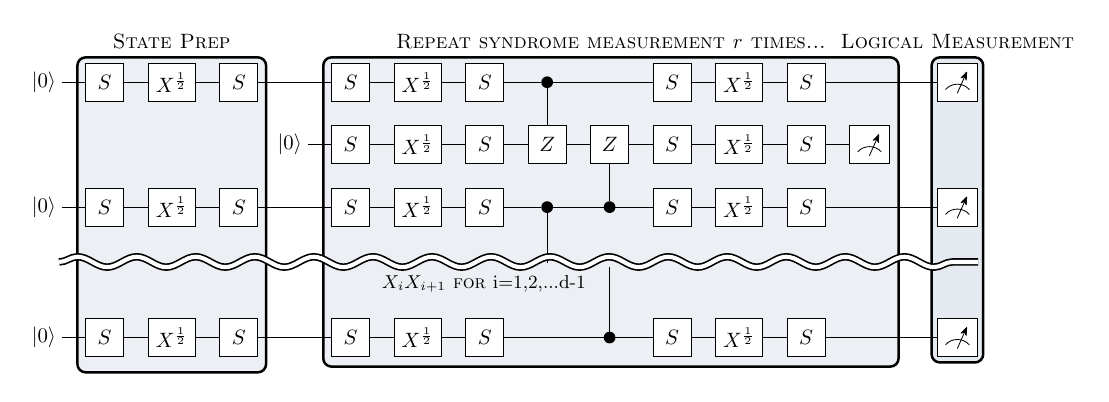}
\caption{Compiled syndrome extraction circuit onto device hardware for $X$-repetition code for a distance $d$ repetition code with $r$ syndrome extraction rounds. State prep in this example prepares the logical $|+\rangle$ state.}
\label{fig:synd_ext_example}
\end{figure*}

\subsection{Training objective and protocol}\label{sec:train}

We optimize binary cross-entropy over per-qubit flip targets, asking only:
\begin{center}
\emph{“Given the detection history and calibration context, what is the probability that data qubit $i$'s target Pauli frame should be XORed?”}
\end{center}
Formally, with predictions $\hat p_{n,i}\in[0,1]$ for shot $n$ and data qubit $i$, we compute the binary cross entropy as:
\begin{equation}
\mathcal{L}
= -\frac{1}{N}\sum_{n=1}^{N}\sum_{i=1}^{d_n}
\big[y_{n,i}\log \hat p_{n,i} + (1-y_{n,i})\log(1-\hat p_{n,i})\big].
\end{equation}

At inference, thresholding $\hat c=\mathbf{1}[\hat p>\tfrac12]$ updates the tracked Pauli frame before computing the logical state.

The network never receives the target logical state, final data-qubit readouts or the logical outcome. Each training example consists only of the detection tensor $\chi\in\{0,1\}^{ r\times(d-1)}$ and the calibration subgraph $G$. Targets $y\in\{0,1\}^d$ are obtained from the observed measurements, and the corresponding corrections that recover the initial logical state.

For each experimental shot $n$ we construct a per–data-qubit flip target $y_n \in \{0,1\}^d$. Let \[q_n = (q_{n,1},\dots,q_{n,d}) \in \{0,1\}^d\] denote the prepared logical bit string, with $q_{n,1} = q_{n,2} =\cdots = q_{n,d}$ for our repetition-code experiments, and let \[m_n = (m_{n,1},\dots,m_{n,d}) \in \{0,1\}^d\] denote the corresponding measured bit string on the data qubits at the final round. We define \[y_n = q_n \oplus m_n,\] where $\oplus$ denotes bitwise XOR. Equivalently, $y_{n,i} = 1$ if the measurement on data qubit $i$ disagrees with the prepared logical value and $y_{n,i} = 0$ otherwise. Thus $y_n$ specifies the Pauli-frame update that, when applied to $m_n$, recovers the prepared logical state $q_n$, and this is the Pauli-frame update that the decoder is trained to predict from the detection history and calibration features.

With respect to training, for each basis ($X$ and $Z$ separately) and each $(d,r)$, we merge shots across IBM \textit{Fez}, \textit{Kingston}, and \textit{Pittsburgh} and split 70\%/30\% for train/validation. We use Adam (initial learning rate $5\times10^{-3}$) with cosine annealing for 100 epochs, validating periodically and checkpointing on validation accuracy improvements. Model parameters are described under Appendix A - Table \ref{tab:nn_params}.

We compare against (i) \textbf{CNN-only} (FiLM disabled; same network backbone, loss, and training) and (ii) a noise-aware minimum weight perfect matching \textbf{MWPM} baseline reported in Sec.~\ref{sec:results}.

\subsection{Complexity of FiLM Decoding Architecture}\label{sec:complexity}

A significant advantage of the FiLM architecture is the complete decoupling of the hardware-conditioning overhead from the real-time decoding loop. As the FiLM parameters $\{\gamma^{(\ell)}, \beta^{(\ell)}\}$ are derived solely from the calibration graph $G$, they do not need to be recomputed for every syndrome tensor $\chi$.

We employ a weight folding strategy, whereby when the convolutional feature map must be updated, the newly generated modulation parameters ($\{\gamma^{(\ell)}, \beta^{(\ell)}\}$) are mapped directly into the respective convolutional weights and biases. This transformation converts the conditional operations into standard convolutions with updated convolutional kernel values. Consequently, the per-shot inference path maintains computational complexity identical to the unconditioned CNN baseline, allowing the decoder to benefit from hardware-aware feature maps with no additional latency penalty during active decoding.

\begin{figure}[htbp]
\centering
\includegraphics{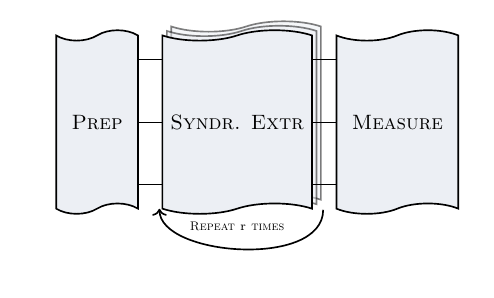}
\caption{Overall repetition code process deployed, where \textsc{PREP} initializes the logical code word, either $\ket{0}_L$ or $\ket{1}_L$ for a $Z-$ repetition code, or $\ket{+}_L$ or $\ket{-}_L$ for an $X-$ repetition code. Syndrome extraction is repeated $r$ times, after which the data qubits are \textsc{Measured}. See Figure \ref{fig:synd_ext_example} for a compiled example over \textsc{IBM Fez} basis gates.}
\label{fig:overall_rep_code}
\end{figure}

\begin{figure}[!t]
  \centering
  \includegraphics[width=0.95\linewidth]{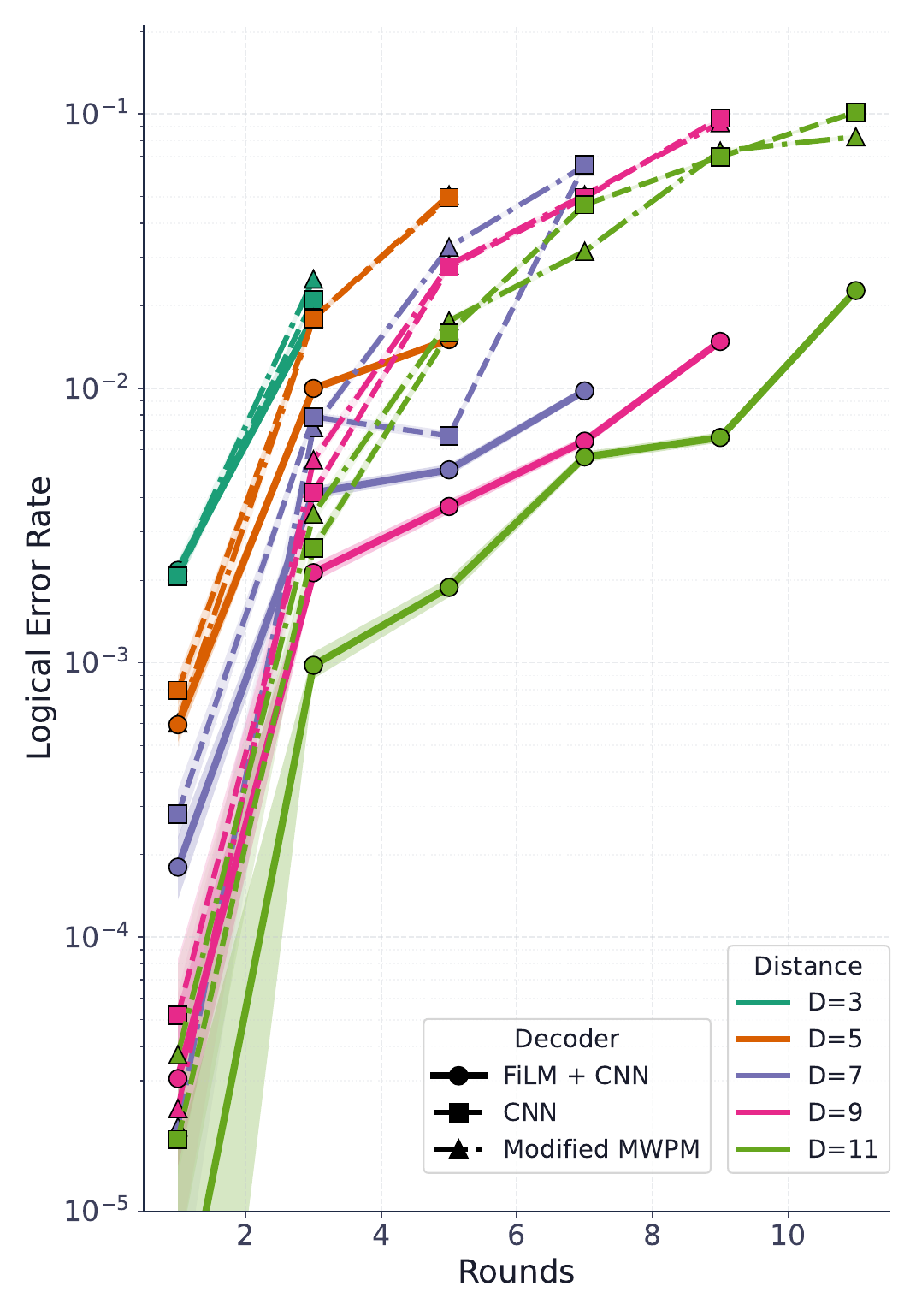}
\caption{Validation results: logical error rates in the X-basis as a function of measurement rounds. Logical states $|+\rangle_L$ and $|-\rangle_L$ are tested. Results are shown for code distances $d={3,5,7,9,11}$ for experiment rounds $r={1,3,..d}$ states. We compare the performance of the proposed FiLM + CNN decoder (solid lines) against a standard CNN (dashed lines) and Modified MWPM (dash-dotted lines). }
  \label{fig:trained_x}
\end{figure}

\begin{figure}[!t]
  \centering
  \includegraphics[width=0.95\linewidth]{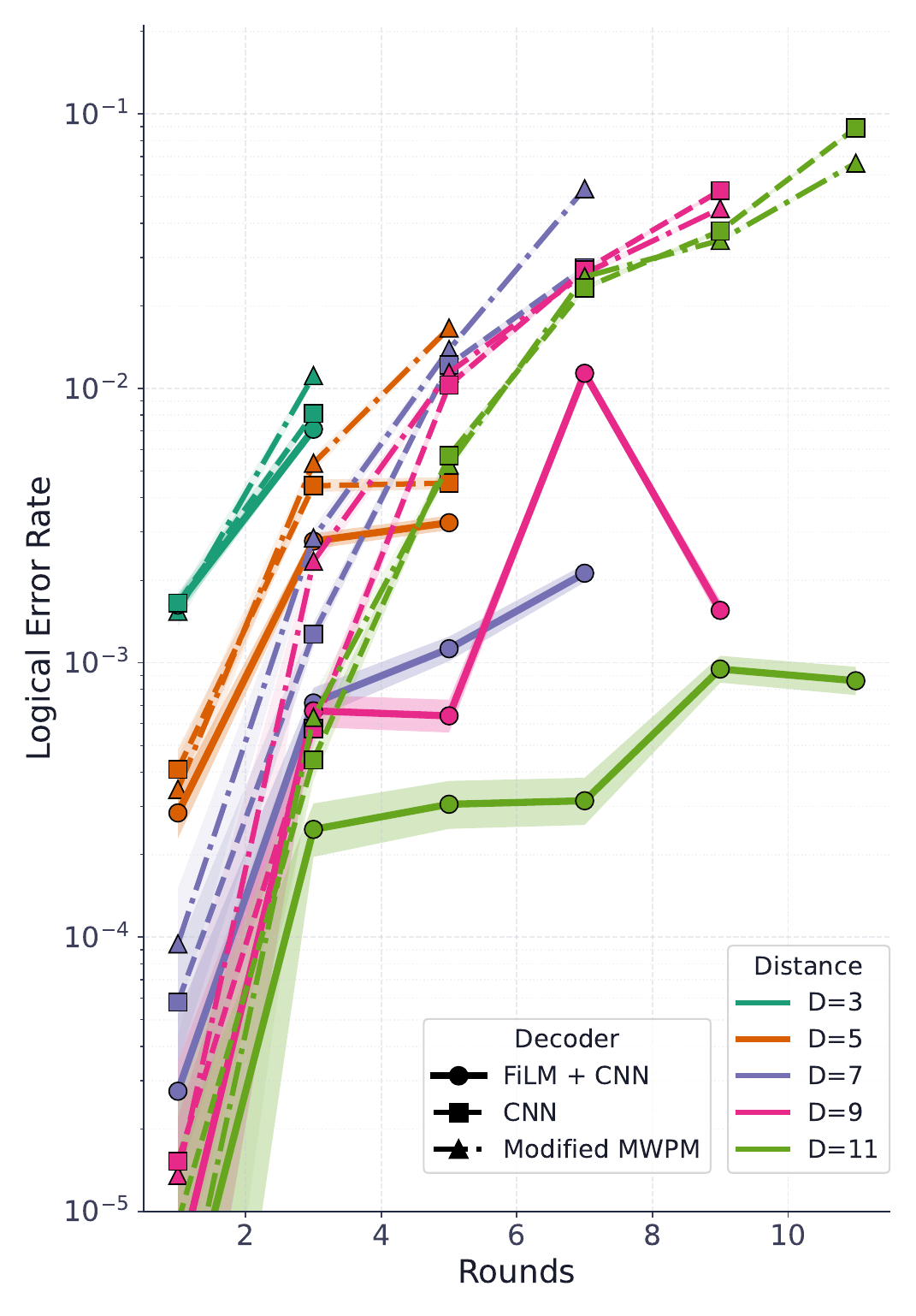}
 \caption{Validation results: logical error rates in the Z-basis as a function of measurement rounds. Logical states $|0\rangle_L$ and $|1\rangle_L$ are tested. Results are shown for code distances $d={3,5,7,9,11}$ for experiment rounds $r={1,3,..d}$ states. We compare the performance of the proposed FiLM + CNN decoder (solid lines) against a standard CNN (dashed lines) and Modified MWPM (dash-dotted lines). }
  \label{fig:trained_z}
\end{figure}

\begin{figure}[!t]
  \centering
  \includegraphics[width=0.95\linewidth]{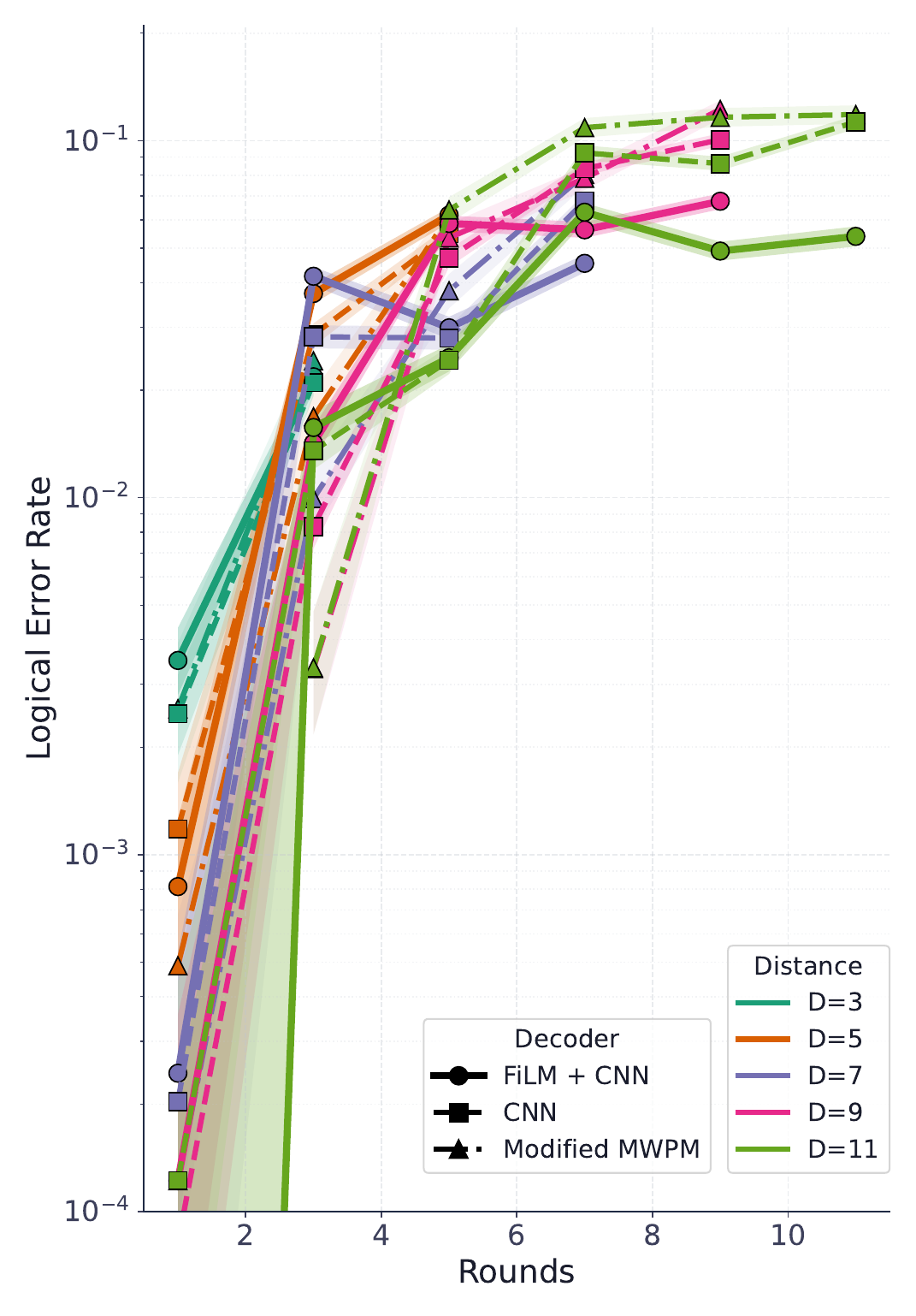}
  \caption{Unseen recalibrated results: logical error rates in the X-basis as a function of measurement rounds for $D\in\{3,5,7,9,11\}$. FiLM\,+\,CNN (solid), CNN-only (dashed), and Modified MWPM (dash-dot). Shaded bands: 95\% CIs across calibration snapshots.}
  \label{fig:overall_x}
\end{figure}

\begin{figure}[!t]
  \centering
  \includegraphics[width=0.95\linewidth]{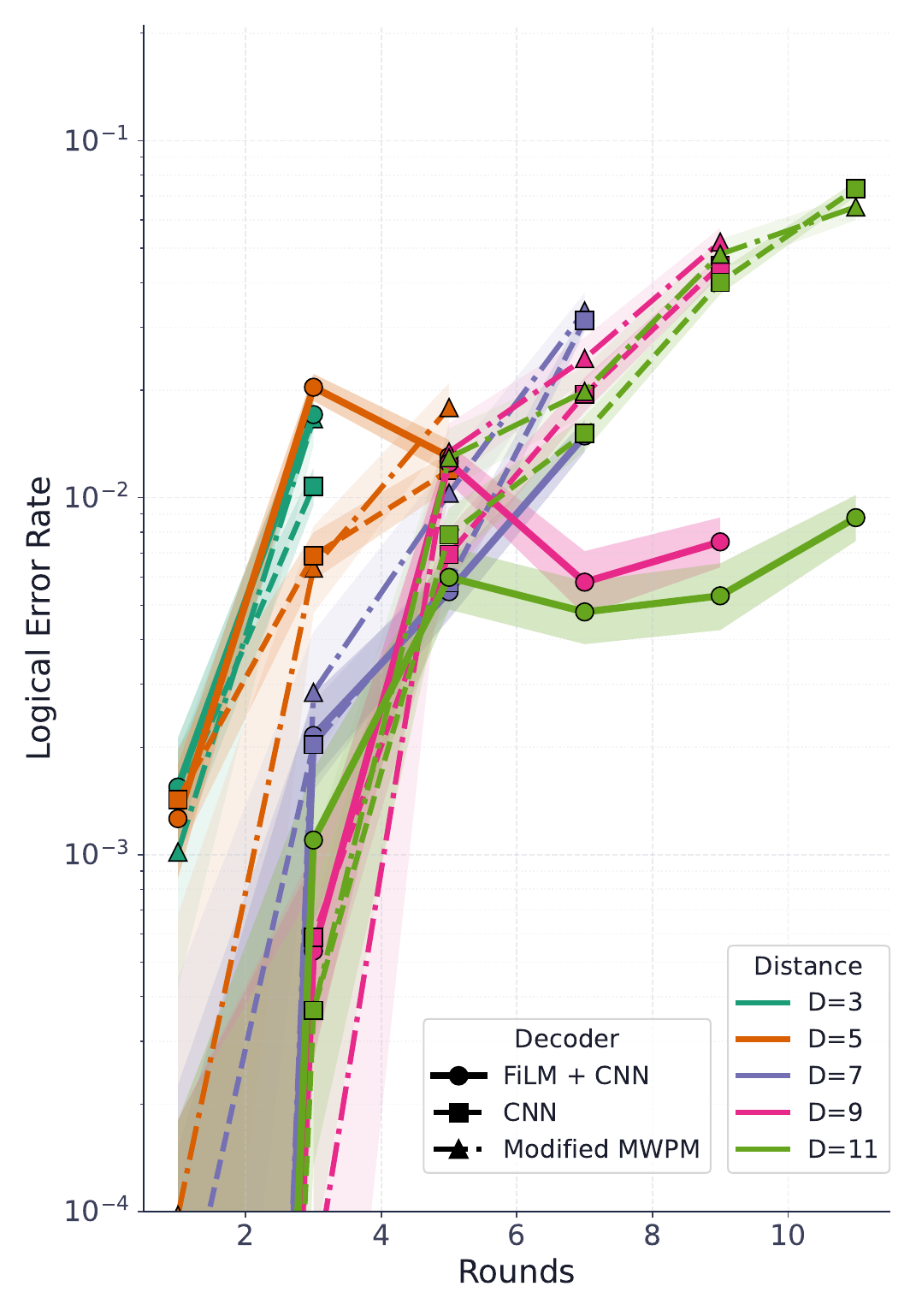}
  \caption{Unseen recalibrated results: logical error rates in the Z-basis matching Fig.~\ref{fig:overall_x}. Separation between FiLM\,+\,CNN and baselines increases at larger $r$ and $d$, consistent with phase-noise bias.}
  \label{fig:overall_z}
\end{figure}

\section{Results}\label{sec:results}

\subsection{Experimental Setup and Evaluation}\label{sec:setup}

We evaluate repetition-code decoders for code distances \(d\in\{3,5,7,9,11\}\) and syndrome depths \(r\in\{1,3,\ldots,d\}\) in both \(X\)- and \(Z\)-basis encodings, using IBM \emph{Fez}, \emph{Kingston}, and \emph{Pittsburgh} processors (heavy-hex connectivity). A total of \(2{,}760{,}704\) shots were collected over \(400\)
contiguous-qubit-chain calibration snapshots. For each \((d,r)\) and basis we train one model on a \(70{:}30\) train/validation split pooled across devices, and then evaluate (i)~on the held-out \textbf{validation data split} and (ii)~on \emph{unseen recalibrated contiguous chains} acquired one week later. As such, all decoding performance reported is on data the models were not trained on. All experiments follow the high level structure depicted in Figure \ref{fig:overall_rep_code}, adjusting the \textsc{PREP}, describing the choice of initial logical state, \textsc{SYNDR. EXTR}, based on the repetition code form, and \textsc{r}, defined as the number of syndrome extraction rounds. Prior to measurement, we map the target observable of each qubit onto the $Z$ basis. An example logical $|+\rangle$ experiment is outlined in Fig \ref{fig:synd_ext_example}, whereby we prepare the $|+\rangle$ state, measure $d-1$ syndromes over $r$ rounds, and finally measure the data qubits. The information attained from the syndromes is decoded and XORed with the data qubits to attain a final logical state measurement.

We compare FiLM+CNN to (i) MWPM and (ii) CNN (identical convolutional backbone with FiLM disabled, trained without FiLM). We report logical error rate (LER), and plot results with binomial 95\% confidence intervals. To quantify relative changes we use:
\[
\ratio{A}{B} \;\equiv\; \frac{\text{LER}_{\!A}}{\text{LER}_{\!B}},
\qquad
\gain{A}{B} \;\equiv\; \frac{1}{\ratio{A}{B}} \;=\; \frac{\text{LER}_{B}}{\text{LER}_{\!A}},
\]
so that \(\ratio{A}{B}<1\) (\(\gain{A}{B}>1\)) indicates that decoder \(A\) improves on \(B\).

\subsection{Validation Data Results}\label{sec:val}

We train one decoder model per $(d,r)$ pair on the 70\% of the training data, and first evaluate logical error rates on the 30\% validation data. Figures~\ref{fig:trained_x}–\ref{fig:trained_z} plot \(\LER\) vs.~\(r\) for all \((d,r)\), with per-configuration values tabulated in Tables~\ref{tab:test_z}–\ref{tab:test_x}. Data points for distance $d$ experiments, over rounds $r \in \{1,3,\dots,d\}$ are plotted. We observe two trends:
\begin{enumerate}
  \item \textbf{Shallow circuits favor \textsc{MWPM}.} For small \(d\) and \(r\), the analytic model in \textsc{MWPM} is competitive or best, especially in the \(X\)-basis.
  \item \textbf{Deeper circuits favor \textsc{FiLM+CNN}.} As \(d\) and \(r\) increase, \textsc{FiLM+CNN}\ overtakes both \textsc{MWPM} and \cnn. At \((d,r)=(11,11)\) we observe a \(\gain{FiLM}{MWPM}=11.11\times\) reduction in \(\LER\) over \textsc{MWPM} in the \(Z\)-basis and a $\gain{FiLM}{MWPM}=5.92\times$ reduction in the dual \(X\)-basis (Tables~\ref{tab:test_z}, \ref{tab:test_x}).
\end{enumerate}

At shallow code depths and few syndrome-extraction rounds, the MWPM decoder retains an advantage. At these depths and small distances, the error graph is sparse. MWPM performs well in these regimes, where underlying noise and error profile matches the generating features for MWPM. Neural networks, conversely, potentially have (1) less data to train on in these regimes due to lower $\chi$ dimensionality, though more importantly, (2) whilst there also being fewer "complex" error patterns to learn as non-Markovian errors and other parasitic correlated noise channels are yet to accumulate. As the circuit volume grows, these assumptions that underpin MWPM start to break down. Temporal and spatial correlations compound, producing error structures that pairwise matching inherently struggles to capture. In the larger-system regime ($d\!\ge\!7,\,r\!\ge\!5$), the learned decoders, and especially the hardware-conditioned \textsc{FiLM+CNN} model, consistently outperform both MWPM and the unconditioned \textsc{CNN}. With respect to the unconditioned \textsc{CNN} model, \textsc{CNN} improves over MWPM in several mid- and large-volume regimes, particularly in the $Z$-basis. However, \textsc{FiLM+CNN} consistently achieves lower LER than \textsc{CNN} across all distances with $r\geq5$ in our experiments on the validation data set, reflecting the benefit of calibration-aware modulation over a purely data-driven but unconditioned convolutional backbone. 

We observe the $Z$-basis configuration achieves a median gain of $4.55\times$ (range $1.99\times$-$11.11\times$) over MWPM in the large-system regime ($d \ge 7, r \ge 5$), whereas the $X$-basis achieves a median gain of $3.19\times$ (range $1.69\times$--$8.77\times$).

The $Z$-basis experiments outperforming the $X$-basis experiments aligns with the noise characteristics of the underlying hardware, where $T_1 \gg T_2$ across the majority of experiments. While the $Z$-repetition code detects bit-flips ($X$ errors) and is theoretically robust to dephasing errors, our FiLM feature analysis (Table \ref{tab:film_svd_x} and \ref{tab:film_svd_z}) indicates that $T_2$ remains a significant driver of feature modulation in the early layers of the $Z$-basis decoder. This sensitivity likely arises from the hardware compilation of the syndrome extraction circuit, whereby the native gate set requires basis-changing operations, such as CZ gates, transiently exposing the $Z$-code to dephasing errors during syndrome extraction. Consequently, the decoder appears to learn that $T_2$ is a relevant predictor of correlated errors even for the $Z$-code. In the $X$-basis, which is continuously sensitive to dephasing errors during idling, the model exhibits a similar pattern, with $T_1$ being a relevant feature. This suggests that the hardware-conditioned architecture attempts to learn the latent dynamics of the circuit implementation, evident by the impact of transient noise during syndrome extraction, rather than relying on an error model based solely on the code's target logical protection. This transient noise exposure can be seen in Figure \ref{fig:synd_ext_example} within the syndrome extraction block.

\subsection{Unseen Recalibrated Experiment Results}\label{sec:unseen}

To verify that the performance gains observed on the validation split are not an artifact of train-validation partitioning, and that the decoder adapts to unseen hardware calibrations, we evaluate the trained decoders on experiments executed on different contiguous qubit chains taken from the \textsc{IBM Kingston} one week later. These chains were selected independent of the training data and have been re-calibrated multiple times, resulting in different hardware noise fingerprints. The decoders are never retrained or fine-tuned between these datasets. Results are plotted and tabulated in Figs.~\ref{fig:overall_x}–\ref{fig:overall_z} and Tables~\ref{tab:val_kingston_x}–\ref{tab:val_kingston_z}.

We observe that the models trained continue to outperform MWPM and \cnn{} in experiments, observing a consistent performance crossover at $\approx (7,5)$ for $Z-$ basis experiments and $\approx (7,7)$ for the dual X- experiments, beyond which the learned decoder consistently outperforms the baselines. At \((d,r)=(11,11)\) we reproduce the strong gains observed on the validation split:
\textsc{FiLM+CNN} reduces LER by \(7.41\times\) in the \(Z\)-basis and \(2.09\times\) in the \(X\)-basis relative
to \textsc{MWPM} (Tables~\ref{tab:val_kingston_z}, \ref{tab:val_kingston_x}).

Performance improvements past $(d,r)$ pairs of \((d\ge 7,\,r\ge 5)\) is observed to be:
\begin{itemize}
  \item \textbf{\(Z\)-basis:} \textsc{FiLM+CNN} improves on \textsc{MWPM} past this regime with a gain of $\gain{FiLM}{MWPM}=1.89\times$ at $(7,5)$, up to $\gain{FiLM}{MWPM}=7.41\times$ at $(11,11)$.
  \item \textbf{\(X\)-basis:} \textsc{FiLM+CNN} improves on \textsc{MWPM} past this regime with a gain of $\gain{FiLM}{MWPM}=1.27\times$ at $(7,5)$, up to $\gain{FiLM}{MWPM}=2.19\times$ at $(11,11)$
\end{itemize}

Comparing to the unconditioned CNN model, we observe that the \textsc{FiLM+CNN} decoder yields up to a $8.33\times$ reduction in LER for the $Z$-basis and up to a $2.09\times$ reduction for the $X$-basis at $(d,r)=(11,11)$ (Tables~\ref{tab:val_kingston_z} and \ref{tab:val_kingston_x}). This highlights the specific value of calibration conditioning in high-noise regimes, where the unconditioned model struggles to generalize.

These results suggest that FiLM’s calibration-driven modulation has learned a form of hardware-adaptive feature selection. The convolutional backbone learns a rich set of reusable spatial–temporal features, while the FiLM parameters selectively amplify or suppress channels depending on the current device calibration. This mechanism helps explain the capability to generalize across days and qubit-chain selections without retraining.

\subsection{Inference Latency}

To quantify the computational overhead of calibration conditioning, we benchmark the per-shot forward-pass latency. We compare three configurations to isolate the cost of the FiLM mechanism:
\begin{enumerate}
    \item \textbf{No-FiLM (CNN):} The baseline convolutional decoder without hardware conditioning.
    \item \textbf{FiLM (Folded):} Calibration-dependent parameters ($\gamma, \beta$) are precomputed and mapped into the convolutional weights.
    \item \textbf{FiLM (Dynamic):} Hardware encoder (GNN) and FiLM generator are executed for every sample.
\end{enumerate}

Benchmarks were conducted using nVidia RTX 5000 GPU. We report the mean and standard deviation over 2,000 decoding iterations following a 500 iteration warmup.

As detailed in Table~\ref{tab:latency}, the \textbf{FiLM (Dynamic)} approach increases latency by over an order of magnitude, resulting in a per-shot latency of $\approx 1.4$\,ms, with execution latency dominated by the graph neural network execution. However, the \textbf{FiLM (Folded)} strategy successfully eliminates this overhead. The folded latencies ($\approx 85\text{--}95\,\mu\text{s}$) track the \textbf{No-FiLM} baseline within the reported measurement error. Notably, latency remains relatively constant as the code distance scales from $d=3$ to $d=11$; in this regime, runtime is dominated by fixed GPU kernel launch overheads rather than computational complexity. This implies that the computational workload is minimal, and on dedicated control hardware such as FPGAs or ASICs, which eliminate kernel scheduling overheads,  we anticipate the latency would drop significantly, governed by the logic and depth of the network \cite{hu2022survey,lee2022energy}. These results highlight that the proposed architecture can embed hardware-conditioning performance gains with next to zero additional latency cost at inference time.

\begin{table}[h]
\centering
\caption{\textbf{Per-Shot Inference Latency (Batch Size = 1).} Comparison of execution time across decoder strategies. \textit{Dynamic} computes FiLM parameters per-shot; \textit{Folded} precomputes them, baking the modulation into the CNN weights. The folded approach incurs negligible overhead compared to the unconditioned baseline.}
\label{tab:latency}
\begin{tabular*}{\columnwidth}{@{\extracolsep{\fill}} l l r r @{}}
\toprule
\textbf{$(d, r)$} & \textbf{Decoder Mode} & \textbf{Avg ($\mu$s)} & \textbf{Std ($\mu$s)} \\
\midrule
\multirow{3}{*}{(3, 3)} 
   & FiLM (Dynamic) & 1419.2 & 280.9 \\
 & FiLM (Folded)  & 96.7   & 27.0  \\
 & No-FiLM (CNN)  & 87.4   & 18.1  \\
\midrule
\multirow{3}{*}{(5, 5)} 
   & FiLM (Dynamic) & 1419.8 & 151.4 \\
 & FiLM (Folded)  & 84.9   & 11.7  \\
 & No-FiLM (CNN)  & 83.3   & 14.5  \\
\midrule
\multirow{3}{*}{(7, 7)} 
   & FiLM (Dynamic) & 1430.1 & 227.0 \\
 & FiLM (Folded)  & 84.4   & 18.5  \\
 & No-FiLM (CNN)  & 84.9   & 16.2  \\
\midrule
\multirow{3}{*}{(9, 9)} 
   & FiLM (Dynamic) & 1426.7 & 203.7 \\
 & FiLM (Folded)  & 93.0   & 17.3  \\
 & No-FiLM (CNN)  & 98.0   & 13.3  \\
\midrule
\multirow{3}{*}{(11, 11)} 
   & FiLM (Dynamic) & 1380.1 & 185.2 \\
 & FiLM (Folded)  & 84.6   & 15.8  \\
 & No-FiLM (CNN)  & 81.1   & 12.4  \\
\bottomrule
\end{tabular*}
\end{table}

\subsection{Interpreting FiLM Features for Decoding}\label{sec:interpret}

To understand how the graph encoder modulates the convolutional backbone, we analyze the sensitivity of the generated FiLM parameters \(\phi = \{\gamma, \beta\}\) to the input calibration features \(\mathbf{f}\). We compute the full experimental dataset averaged Jacobian \(\mathbf{J} = \mathbb{E}[\partial \phi / \partial \mathbf{f}]\) and perform a Singular Value Decomposition (SVD) to extract the dominant modes of the Jacobian.

Detailed statistics are reported in Table~\ref{tab:film_svd_z} (\(Z\)-basis) and Table~\ref{tab:film_svd_x} (\(X\)-basis). These tables present the singular values (\(\sigma\)) indicating the mode's strength and the normalized projection of the singular vector onto specific hardware features (\(T_1, T_2, \epsilon_g, \epsilon_r\)). Focusing on the largest $(d,r)$ size \((d=11)\), we observe three distinct features:
\begin{itemize}
    \item \textbf{Physical Feature Extraction (Layer 1):} 
    Across both experiments, the first convolutional layer acts as a physical feature extractor, modulating filters based on mixtures of coherence parameters or gate and measurement errors. In the X-basis (Table~\ref{tab:film_svd_x}), the dominant mode ($\gamma=1.81, \beta=3.18$) exhibits a coupled sensitivity, driven primarily by Dephasing ($T_2$, $\gamma=+0.81, \beta=+0.83$) but significantly counter-weighted by Relaxation ($T_1$, $\gamma=-0.58, \beta=-0.54$). Similarly, in the Z-basis (Table~\ref{tab:film_svd_z}), the encoder identifies both coherence channels as prevalent features. While the first mode isolates Readout error, subsequent modes explicitly target $T_1$ (Mode 2) and $T_2$ (Mode 3). This provides evidence that Layer 1 utilizes the full calibration profile ($T_1$ and $T_2$) to inform decision-making, irrespective of code type, rather than relying on a single dominant error channel.
    
    \item \textbf{Partitioning of Measurement Noise (Layer 1):}
    The Z-basis analysis reveals a seemingly particular separation of error sources in the first layer. Mode 1 is driven primarily by Readout Error ($\epsilon_r$, weight $\gamma=\mathbf{+0.95}, \beta=\mathbf{-0.95}$), while Mode 2 captures Relaxation ($T_1$, weight $\gamma=\mathbf{+0.96}, \beta=\mathbf{+0.98}$). This is consistent as well in the X-basis analysis, whereby the second mode is dedicated almost exclusively to Readout Error ($\gamma=\mathbf{+0.95}, \beta=\mathbf{+0.96}$). This provides evidence that the network learns to semantically split measurement uncertainty from data qubit errors in the first convolutional layer.

    \item \textbf{Transition from specific feature modulation to learned kernel modulation (L2/L3):} 
    In deeper layers, the association with specific error channels seemingly diminishes. By Layer 3 of the X-basis model, the dominant mode ($\gamma=0.53, \beta=4.01$) exhibits mixed sensitivity to noise model features, coupling $T_2$ ($\gamma=+0.76, \beta=+0.74$), $T_1$ ($\gamma=-0.42, \beta=-0.40$), and Readout errors ($\gamma=+0.47, \beta=+0.53$) simultaneously. A similar pattern emerges in the $Z-$ basis experiments, whereby the top mode ($\gamma=0.65, \beta=3.45$) couples to a mixture of $T_1$ ($\gamma=-0.41, \beta=-0.04$), $T_2$ ($\gamma=+0.41, \beta=-0.25$), and Readout errors ($\gamma=+0.81, \beta=-0.97$). This evolution motivates our hypothesis that the network hierarchy is learning the decoding task process. Layer 1 performs calibration-aware feature extraction (isolating physical error sources), while deeper layers perform the logical inference, re-weighting these extracted feature maps to resolve particular detector patterns where the distinction between a measurement error and a data qubit error is context-dependent. 
\end{itemize}

The Jacobian analyses suggest that FiLM acts as a soft attention mechanism, whereby the weights are continuous features generated via a learned hardware embedding. Rather than learning a static decoding rule, the network learns a library of filters and a dynamic policy for deploying them. The hardware encoder provides decoding context based on prior extracted hardware features from which the model learns underlying correlations. This mechanism helps explain the model's ability to generalize to unseen chains without retraining, as the convolutional weights provide the decoding capability, while the calibration embedding provides the decoding context.

\section{Discussion}\label{sec:discussion}

The spatiotemporal variability of quantum hardware typically forces a trade-off between decoder accuracy and the operational cost in either pretraining or induction. In this work, we put forward a calibration-conditioned neural decoder as a strategy to mitigate this challenge. Our experimental results suggest that conditioning on a learned embedding of calibration data allows the decoder to maintain performance across hardware changes, with promising results in the domain of adaptive decoding. However, we acknowledge the results are limited to the 1D repetition code, and hence as hardware evolves further tests on larger codes is needed.

\subsection{Generalization via Calibration Conditioning}

An observation of our work is the decoder's continued performant capabilities on unseen qubit chains and recalibrated devices. The ability to function on new hardware configurations suggests that the model is not strictly memorizing the spatial locations of erroneous qubits or specific recurring error channels from the training set. Instead, the results are consistent with the network learning a generalized conditional probability distribution within the calibration embedding space. The model appears to learn how specific calibrations, such as a drop in readout fidelity or an increase in $T_1$ relaxation, correlate with error likelihoods, regardless of their specific location.

Whilst the network only is explicitly provided standard calibration metrics that are used to derive error rates for MWPM ($T_1$,$T_2$, gate errors), the performance gains suggest the GCN encoder may also be inferring latent noise properties of the system based on the learned correlation between calibration data, topology information and experiment data. This capability is particularly important for fault-tolerant computation where logical qubits are not static entities. Protocols such as lattice surgery or code deformations inherently require decoders to dynamically shift their physical support during computation. As such, maintaining a bespoke trained decoder for every physical qubit permutation is unreasonable. A single, adaptable decoder that generalises sufficiently across hardware provides the necessary flexibility for these dynamic operations, effectively decoupling the decoding policy from specific physical qubits.

\subsection{Latency and Implementation}

From an operational standpoint, the proposed architecture demonstrates that a systems noise model could be integrated potentially without incurring a latency penalty. Our empirical benchmarks demonstrate that by folding the pre-computed FiLM parameters ($\gamma, \beta$) directly into the convolutional weights, the real-time inference complexity becomes approximately identical to the unconditioned baseline. This effectively decouples the computational cost of the hardware encoder from the critical low-latency path of the decoder. 

This neural architecture exploits the natural separation of timescales in superconducting quantum processors. Calibration data shifts occur over hours or days, whereas syndrome extraction occurs at the microsecond level. By performing the heavy-lifting of the graph neural network asynchronously, the real-time decoding path remains lightweight. 

\section{Conclusion and Future Work}\label{sec:conclusion}

In this work, we introduced a hardware-conditioned neural decoder that leverages Feature-wise Linear Modulation (FiLM) to adapt dynamically to the spatiotemporal variability of superconducting quantum processors. By evaluating over 2.7 million experimental shots on IBM devices, we observed that this approach yields lower logical error rates compared to matching-based and unconditioned neural baselines, particularly in large-distance regimes ($d \geq 7$) where error correlations are most pronounced. Furthermore, we demonstrate that a single model is capable of generalising to unseen contiguous qubit chains and future calibration snapshots without retraining, suggesting the decoder can learn a transferable representation of hardware noise. 

A natural extension and future work is to extend this conditioning framework to higher-weight stabilizer codes, such as the surface code, where the calibration graph naturally generalizes to support dual $X$- and $Z$-basis decoding. We anticipate that the latency benefits of the architecture will persist, as the heavy-lifting of the graph neural network is performed via pre-computable FiLM parameters. A primary challenge lies in designing topological backbones capable of navigating more complex code stabilizer graphs while retaining the efficiency of the conditioned convolutional approach.

\section*{Acknowledgements}
This material is based upon work supported by the U.S. Department of Energy, Office of Science, National Quantum Information Science Research Centers, Quantum Science Center (QSC). This research was supported by PNNL’s Quantum Algorithms and Architecture for Domain Science (QuAADS) Laboratory Directed Research and Development (LDRD) Initiative.  The Pacific Northwest National Laboratory is operated by Battelle for the U.S. Department of Energy under Contract DE-AC05-76RL01830. This research used resources of the Oak Ridge Leadership Computing Facility (OLCF), which is a DOE Office of Science User Facility supported under Contract DE-AC05-00OR22725. This research used resources of the National Energy Research Scientific Computing Center (NERSC), a U.S. Department of Energy Office of Science User Facility located at Lawrence Berkeley National Laboratory, operated under Contract No. DE-AC02-05CH11231. 

\newpage

\bibliography{refs}

\newpage
\appendix
\section*{Appendix}

\section{Neural Network Parameters}
The FiLM-conditioned decoder couples a three-layer 3\,$\times$\,3 CNN with FiLM scalings
derived from a three-layer GCN over hardware graphs.
\begin{table}[h]
  \centering
  \small
  \setlength{\tabcolsep}{4pt}
  \renewcommand{\arraystretch}{1.2}
  \begin{tabular}{@{} l l @{}}
  \toprule
  \textbf{Parameter} & \textbf{Value} \\
  \midrule
  \multicolumn{2}{l}{\textit{Architecture}} \\
  Main Network Arch. & 2D Convolutional (3 blocks) \\
  Conv Kernel Config & $k=3$, padding $=1$, ReLU \\
  Channel Depth & $128 \to 256 \to 512$ \\
  Hardware Encoder & 3-layer GCN \\
  Latent Dimension & 256 \\
  FiLM Generator & MLP (256 $\to$ 256 $\to$ Out) \\
  Output Head & Sigmoid \\
  \midrule
  \multicolumn{2}{l}{\textit{Training}} \\
  Batch Size & 4096 \\
  Epochs & 100 \\
  Data Split & 70\% Train / 30\% Val \\
  Optimizer & Adam ($\beta_1=0.9, \beta_2=0.999$) \\
  Learning Rate & $5 \times 10^{-3}$ \\
  Scheduler & Cosine Annealing ($\eta_{\min}=0$) \\
  Loss Function & Binary Cross-Entropy \\
  \bottomrule
  \end{tabular}
  \caption{FiLM+CNN decoder architecture and training summary.}
  \label{tab:nn_params}
\end{table}

\subsection{Logical Error Rates}
Tables~\ref{tab:test_z} through \ref{tab:val_kingston_x} present the detailed breakdown of logical error rates (LER) across all experimental configurations. Tables~\ref{tab:test_z} and \ref{tab:test_x} correspond to the unseen validation set, while Tables~\ref{tab:val_kingston_z} and \ref{tab:val_kingston_x} correspond to the unseen data on the Kingston device, sampled one week later. Values of '—' and $0.00$ indicate regimes where no logical errors were observed for the baseline decoder, rendering ratio comparisons undefined. These configurations correspond to shallow circuits where all decoders succeed post majority vote.
\begingroup
\setlength{\tabcolsep}{6pt} 
\renewcommand{\arraystretch}{1.1} 

\begin{table*}[t]
  \centering
  \small
  \caption{Z-Basis Decoder comparison on \textbf{Validation Set}. Comparison of Logical Error Rates (LER) for FiLM+CNN, Modified MWPM, and Unconditioned CNN. The Ratio $\rho = \text{LER}_{\text{FiLM}} / \text{LER}_{\text{Base}}$ indicates performance gain (values $<1.0$ indicate FiLM improvement).  Values of '—' and $0.00$ no logical errors.}
  \label{tab:test_z}
  \begin{tabular}{cc ccc cc}
    \toprule
    \multicolumn{2}{c}{Setup} & \multicolumn{3}{c}{Logical Error Rate (LER)} & \multicolumn{2}{c}{Ratio ($\rho$)} \\
    \cmidrule(lr){1-2} \cmidrule(lr){3-5} \cmidrule(lr){6-7}
    $d$ & $r$ & FiLM+CNN & MWPM & CNN & vs MWPM & vs CNN \\
    \midrule
    3 & 1 & $1.30 \times 10^{-3}$ & $1.54 \times 10^{-3}$ & $1.07 \times 10^{-3}$ & \textbf{0.845} & 1.22 \\
    3 & 3 & $1.16 \times 10^{-2}$ & $1.11 \times 10^{-2}$ & $1.10 \times 10^{-2}$ & 1.05 & 1.06 \\
    \addlinespace
    5 & 1 & $7.52 \times 10^{-4}$ & $3.46 \times 10^{-4}$ & $7.76 \times 10^{-4}$ & 2.18 & 0.970 \\
    5 & 3 & $1.33 \times 10^{-2}$ & $5.34 \times 10^{-3}$ & $1.02 \times 10^{-2}$ & 2.48 & 1.29 \\
    5 & 5 & $1.72 \times 10^{-2}$ & $1.66 \times 10^{-2}$ & $1.58 \times 10^{-2}$ & 1.04 & 1.09 \\
    \addlinespace
    7 & 1 & $7.57 \times 10^{-5}$ & $9.46 \times 10^{-5}$ & $7.57 \times 10^{-5}$ & \textbf{0.800} & 1.00 \\
    7 & 3 & $2.59 \times 10^{-3}$ & $2.85 \times 10^{-3}$ & $2.09 \times 10^{-3}$ & \textbf{0.908} & 1.24 \\
    7 & 5 & $6.08 \times 10^{-3}$ & $1.39 \times 10^{-2}$ & $1.36 \times 10^{-2}$ & \textbf{0.438} & \textbf{0.447} \\
    7 & 7 & $1.19 \times 10^{-2}$ & $5.34 \times 10^{-2}$ & $3.18 \times 10^{-2}$ & \textbf{0.222} & \textbf{0.373} \\
    \addlinespace
    9 & 1 & $0.00$ & $1.36 \times 10^{-5}$ & $0.00$ & \textbf{0.000} & --- \\
    9 & 3 & $9.03 \times 10^{-4}$ & $2.35 \times 10^{-3}$ & $9.77 \times 10^{-4}$ & \textbf{0.385} & \textbf{0.925} \\
    9 & 5 & $3.25 \times 10^{-3}$ & $1.14 \times 10^{-2}$ & $6.59 \times 10^{-3}$ & \textbf{0.284} & \textbf{0.493} \\
    9 & 7 & $3.42 \times 10^{-3}$ & $2.62 \times 10^{-2}$ & $3.48 \times 10^{-2}$ & \textbf{0.130} & \textbf{0.098} \\
    9 & 9 & $9.94 \times 10^{-3}$ & $4.53 \times 10^{-2}$ & $2.54 \times 10^{-2}$ & \textbf{0.220} & \textbf{0.392} \\
    \addlinespace
    11 & 1 & $0.00$ & $3.21 \times 10^{-6}$ & $0.00$ & \textbf{0.000} & --- \\
    11 & 3 & $7.73 \times 10^{-4}$ & $6.33 \times 10^{-4}$ & $2.85 \times 10^{-4}$ & 1.22 & 2.71 \\
    11 & 5 & $2.65 \times 10^{-3}$ & $5.26 \times 10^{-3}$ & $5.09 \times 10^{-3}$ & \textbf{0.503} & \textbf{0.520} \\
    11 & 7 & $2.93 \times 10^{-3}$ & $2.55 \times 10^{-2}$ & $1.87 \times 10^{-2}$ & \textbf{0.115} & \textbf{0.157} \\
    11 & 9 & $5.17 \times 10^{-3}$ & $3.47 \times 10^{-2}$ & $4.98 \times 10^{-2}$ & \textbf{0.149} & \textbf{0.104} \\
    11 & 11 & $5.96 \times 10^{-3}$ & $6.63 \times 10^{-2}$ & $9.42 \times 10^{-2}$ & \textbf{0.090} & \textbf{0.063} \\
    \bottomrule
  \end{tabular}
\end{table*}

\begin{table*}[t]
  \centering
  \small
  \caption{X-Basis Decoder comparison on \textbf{Validation Set}. Comparison of Logical Error Rates (LER) for FiLM+CNN, Modified MWPM, and Unconditioned CNN. The Ratio $\rho = \text{LER}_{\text{FiLM}} / \text{LER}_{\text{Base}}$ indicates performance gain (values $<1.0$ indicate FiLM improvement). Values of '—' and $0.00$ no logical errors.}
  \label{tab:test_x}
  \begin{tabular}{cc ccc cc}
    \toprule
    \multicolumn{2}{c}{Setup} & \multicolumn{3}{c}{Logical Error Rate (LER)} & \multicolumn{2}{c}{Ratio ($\rho$)} \\
    \cmidrule(lr){1-2} \cmidrule(lr){3-5} \cmidrule(lr){6-7}
    $d$ & $r$ & FiLM+CNN & MWPM & CNN & vs MWPM & vs CNN \\
    \midrule
    3 & 1 & $2.76 \times 10^{-3}$ & $2.14 \times 10^{-3}$ & $2.33 \times 10^{-3}$ & 1.29 & 1.18 \\
    3 & 3 & $2.28 \times 10^{-2}$ & $2.50 \times 10^{-2}$ & $2.28 \times 10^{-2}$ & \textbf{0.912} & 1.00 \\
    \addlinespace
    5 & 1 & $8.79 \times 10^{-4}$ & $6.04 \times 10^{-4}$ & $6.59 \times 10^{-4}$ & 1.46 & 1.33 \\
    5 & 3 & $1.62 \times 10^{-2}$ & $1.80 \times 10^{-2}$ & $9.89 \times 10^{-3}$ & \textbf{0.897} & 1.64 \\
    5 & 5 & $1.96 \times 10^{-2}$ & $5.08 \times 10^{-2}$ & $2.38 \times 10^{-2}$ & \textbf{0.386} & \textbf{0.823} \\
    \addlinespace
    7 & 1 & $1.53 \times 10^{-4}$ & $2.04 \times 10^{-5}$ & $1.22 \times 10^{-4}$ & 7.50 & 1.25 \\
    7 & 3 & $1.07 \times 10^{-2}$ & $7.21 \times 10^{-3}$ & $3.35 \times 10^{-3}$ & 1.49 & 3.21 \\
    7 & 5 & $1.03 \times 10^{-2}$ & $3.28 \times 10^{-2}$ & $1.19 \times 10^{-2}$ & \textbf{0.313} & \textbf{0.862} \\
    7 & 7 & $3.84 \times 10^{-2}$ & $6.47 \times 10^{-2}$ & $2.87 \times 10^{-2}$ & \textbf{0.593} & 1.34 \\
    \addlinespace
    9 & 1 & $4.07 \times 10^{-5}$ & $2.37 \times 10^{-5}$ & $0.00$ & 1.71 & --- \\
    9 & 3 & $1.00 \times 10^{-2}$ & $5.49 \times 10^{-3}$ & $9.03 \times 10^{-3}$ & 1.82 & 1.11 \\
    9 & 5 & $1.04 \times 10^{-2}$ & $2.82 \times 10^{-2}$ & $1.40 \times 10^{-2}$ & \textbf{0.367} & \textbf{0.741} \\
    9 & 7 & $2.37 \times 10^{-2}$ & $5.07 \times 10^{-2}$ & $2.08 \times 10^{-2}$ & \textbf{0.467} & 1.14 \\
    9 & 9 & $1.83 \times 10^{-2}$ & $9.30 \times 10^{-2}$ & $4.96 \times 10^{-2}$ & \textbf{0.197} & \textbf{0.368} \\
    \addlinespace
    11 & 1 & $0.00$ & $3.73 \times 10^{-5}$ & $0.00$ & \textbf{0.000} & --- \\
    11 & 3 & $3.17 \times 10^{-3}$ & $3.49 \times 10^{-3}$ & $1.22 \times 10^{-3}$ & \textbf{0.910} & 2.60 \\
    11 & 5 & $3.54 \times 10^{-3}$ & $1.76 \times 10^{-2}$ & $6.78 \times 10^{-3}$ & \textbf{0.201} & \textbf{0.523} \\
    11 & 7 & $3.60 \times 10^{-3}$ & $3.16 \times 10^{-2}$ & $1.67 \times 10^{-2}$ & \textbf{0.114} & \textbf{0.215} \\
    11 & 9 & $4.32 \times 10^{-2}$ & $7.33 \times 10^{-2}$ & $4.70 \times 10^{-2}$ & \textbf{0.590} & \textbf{0.919} \\
    11 & 11 & $1.40 \times 10^{-2}$ & $8.26 \times 10^{-2}$ & $3.63 \times 10^{-2}$ & \textbf{0.169} & \textbf{0.385} \\
    \bottomrule
  \end{tabular}
\end{table*}

\begin{table*}[t]
  \centering
  \small
  \caption{Z-Basis Decoder performance on \textbf{Kingston} unseen experimental data. Experiments are run one week later, on unseen contiguous qubit chains. Values of '—' and $0.00$ no logical errors.}
  \label{tab:val_kingston_z}
  \begin{tabular}{cc ccc cc}
    \toprule
    \multicolumn{2}{c}{Setup} & \multicolumn{3}{c}{Logical Error Rate (LER)} & \multicolumn{2}{c}{Ratio ($\rho$)} \\
    \cmidrule(lr){1-2} \cmidrule(lr){3-5} \cmidrule(lr){6-7}
    $d$ & $r$ & FiLM+CNN & MWPM & CNN & vs MWPM & vs CNN \\
    \midrule
    3 & 1 & $1.55 \times 10^{-3}$ & $1.02 \times 10^{-3}$ & $1.42 \times 10^{-3}$ & 1.52 & 1.09 \\
    3 & 3 & $1.71 \times 10^{-2}$ & $1.66 \times 10^{-2}$ & $1.12 \times 10^{-2}$ & 1.03 & 1.53 \\
    \addlinespace
    5 & 1 & $1.26 \times 10^{-3}$ & $9.77 \times 10^{-5}$ & $1.30 \times 10^{-3}$ & 12.9 & 0.969 \\
    5 & 3 & $2.04 \times 10^{-2}$ & $6.35 \times 10^{-3}$ & $6.80 \times 10^{-3}$ & 3.21 & 3.00 \\
    5 & 5 & $1.30 \times 10^{-2}$ & $1.79 \times 10^{-2}$ & $9.97 \times 10^{-3}$ & \textbf{0.728} & 1.31 \\
    \addlinespace
    7 & 1 & $0.00$ & $0.00$ & $4.07 \times 10^{-5}$ & --- & \textbf{0.000} \\
    7 & 3 & $2.16 \times 10^{-3}$ & $2.85 \times 10^{-3}$ & $2.04 \times 10^{-3}$ & \textbf{0.757} & 1.06 \\
    7 & 5 & $5.45 \times 10^{-3}$ & $1.03 \times 10^{-2}$ & $5.78 \times 10^{-3}$ & \textbf{0.529} & \textbf{0.944} \\
    7 & 7 & $1.49 \times 10^{-2}$ & $3.33 \times 10^{-2}$ & $3.13 \times 10^{-2}$ & \textbf{0.446} & \textbf{0.474} \\
    \addlinespace
    9 & 1 & $0.00$ & $0.00$ & $0.00$ & --- & --- \\
    9 & 3 & $5.37 \times 10^{-4}$ & $6.10 \times 10^{-5}$ & $4.40 \times 10^{-4}$ & 8.80 & 1.22 \\
    9 & 5 & $1.25 \times 10^{-2}$ & $1.34 \times 10^{-2}$ & $6.93 \times 10^{-3}$ & \textbf{0.931} & 1.80 \\
    9 & 7 & $5.80 \times 10^{-3}$ & $2.45 \times 10^{-2}$ & $1.95 \times 10^{-2}$ & \textbf{0.236} & \textbf{0.298} \\
    9 & 9 & $7.52 \times 10^{-3}$ & $5.20 \times 10^{-2}$ & $4.47 \times 10^{-2}$ & \textbf{0.145} & \textbf{0.168} \\
    \addlinespace
    11 & 1 & $0.00$ & $0.00$ & $0.00$ & --- & --- \\
    11 & 3 & $1.10 \times 10^{-3}$ & $3.66 \times 10^{-4}$ & $1.71 \times 10^{-3}$ & 3.00 & \textbf{0.643} \\
    11 & 5 & $5.98 \times 10^{-3}$ & $1.29 \times 10^{-2}$ & $7.87 \times 10^{-3}$ & \textbf{0.462} & \textbf{0.760} \\
    11 & 7 & $4.79 \times 10^{-3}$ & $1.98 \times 10^{-2}$ & $1.51 \times 10^{-2}$ & \textbf{0.241} & \textbf{0.316} \\
    11 & 9 & $5.31 \times 10^{-3}$ & $4.81 \times 10^{-2}$ & $4.01 \times 10^{-2}$ & \textbf{0.110} & \textbf{0.132} \\
    11 & 11 & $8.79 \times 10^{-3}$ & $6.52 \times 10^{-2}$ & $7.33 \times 10^{-2}$ & \textbf{0.135} & \textbf{0.120} \\
    \bottomrule
  \end{tabular}
\end{table*}

\begin{table*}[t]
  \centering
  \small
  \caption{X-Basis Decoder performance on \textbf{Kingston} unseen experimental data. Experiments are run one week later, on unseen contiguous qubit chains. Values of '—' and $0.00$ no logical errors.}
  \label{tab:val_kingston_x}
  \begin{tabular}{cc ccc cc}
    \toprule
    \multicolumn{2}{c}{Setup} & \multicolumn{3}{c}{Logical Error Rate (LER)} & \multicolumn{2}{c}{Ratio ($\rho$)} \\
    \cmidrule(lr){1-2} \cmidrule(lr){3-5} \cmidrule(lr){6-7}
    $d$ & $r$ & FiLM+CNN & MWPM & CNN & vs MWPM & vs CNN \\
    \midrule
    3 & 1 & $3.50 \times 10^{-3}$ & $2.56 \times 10^{-3}$ & $2.48 \times 10^{-3}$ & 1.37 & 1.41 \\
    3 & 3 & $2.18 \times 10^{-2}$ & $2.42 \times 10^{-2}$ & $1.91 \times 10^{-2}$ & \textbf{0.902} & 1.14 \\
    \addlinespace
    5 & 1 & $8.14 \times 10^{-4}$ & $4.88 \times 10^{-4}$ & $1.30 \times 10^{-3}$ & 1.67 & \textbf{0.625} \\
    5 & 3 & $3.73 \times 10^{-2}$ & $1.69 \times 10^{-2}$ & $3.33 \times 10^{-2}$ & 2.22 & 1.12 \\
    5 & 5 & $6.19 \times 10^{-2}$ & $6.14 \times 10^{-2}$ & $4.30 \times 10^{-2}$ & 1.01 & 1.44 \\
    \addlinespace
    7 & 1 & $2.44 \times 10^{-4}$ & $1.22 \times 10^{-4}$ & $2.44 \times 10^{-4}$ & 2.00 & 1.00 \\
    7 & 3 & $4.17 \times 10^{-2}$ & $9.95 \times 10^{-3}$ & $2.82 \times 10^{-2}$ & 4.19 & 1.48 \\
    7 & 5 & $2.99 \times 10^{-2}$ & $3.80 \times 10^{-2}$ & $2.02 \times 10^{-2}$ & \textbf{0.787} & 1.48 \\
    7 & 7 & $4.53 \times 10^{-2}$ & $8.04 \times 10^{-2}$ & $6.77 \times 10^{-2}$ & \textbf{0.563} & \textbf{0.669} \\
    \addlinespace
    9 & 1 & $1.22 \times 10^{-4}$ & $0.00$ & $8.14 \times 10^{-5}$ & --- & 1.50 \\
    9 & 3 & $1.42 \times 10^{-2}$ & $3.32 \times 10^{-3}$ & $6.63 \times 10^{-3}$ & 4.28 & 2.14 \\
    9 & 5 & $5.86 \times 10^{-2}$ & $5.34 \times 10^{-2}$ & $4.69 \times 10^{-2}$ & 1.10 & 1.25 \\
    9 & 7 & $5.63 \times 10^{-2}$ & $7.85 \times 10^{-2}$ & $8.35 \times 10^{-2}$ & \textbf{0.717} & \textbf{0.674} \\
    9 & 9 & $6.77 \times 10^{-2}$ & $1.22 \times 10^{-1}$ & $1.01 \times 10^{-1}$ & \textbf{0.554} & \textbf{0.672} \\
    \addlinespace
    11 & 1 & $0.00$ & $0.00$ & $1.22 \times 10^{-4}$ & --- & \textbf{0.000} \\
    11 & 3 & $1.57 \times 10^{-2}$ & $3.34 \times 10^{-3}$ & $1.09 \times 10^{-2}$ & 4.71 & 1.44 \\
    11 & 5 & $2.47 \times 10^{-2}$ & $6.40 \times 10^{-2}$ & $2.43 \times 10^{-2}$ & \textbf{0.385} & 1.02 \\
    11 & 7 & $6.31 \times 10^{-2}$ & $1.09 \times 10^{-1}$ & $9.25 \times 10^{-2}$ & \textbf{0.579} & \textbf{0.682} \\
    11 & 9 & $4.91 \times 10^{-2}$ & $1.16 \times 10^{-1}$ & $8.62 \times 10^{-2}$ & \textbf{0.422} & \textbf{0.569} \\
    11 & 11 & $5.40 \times 10^{-2}$ & $1.18 \times 10^{-1}$ & $1.13 \times 10^{-1}$ & \textbf{0.456} & \textbf{0.478} \\
    \bottomrule
  \end{tabular}
\end{table*}
\endgroup

\subsection{Interpretation of Learned FiLM Modes}

To investigate the physical mechanism of the decoder, we perform a Singular Value Decomposition (SVD) on the Jacobian of the FiLM generator. Table~\ref{tab:film_svd_z} (Z-basis) and Table~\ref{tab:film_svd_x} (X-basis) detail the dominant singular modes.

We decompose the sensitivity of the FiLM parameters ($\theta \in \{\gamma, \beta\}$) to the input hardware calibration features. The table separates the normalized contributions of Relaxation ($T_1$), Dephasing ($T_2$), Gate Error ($\epsilon_g$), and Readout Assignment Error ($\epsilon_r$) under features ($\mathbf{f}$). Dominant drivers for each mode are highlighted in \textbf{bold}.

\begingroup
\setlength{\tabcolsep}{4pt} 
\renewcommand{\arraystretch}{1.15} 
\begingroup
\setlength{\tabcolsep}{12pt} 
\renewcommand{\arraystretch}{1.15} 

\begin{table*}[t]
  \centering
  \small
  \caption{\textbf{X-Basis Dominant FiLM Modes ($d=11$).} The Feature columns show the sensitivity of the mode to specific hardware defects. Values represent the normalized singular vector component (signed), with dominant drivers highlighted in bold.}
  \label{tab:film_svd_x}
  
  \begin{tabular}{cc cc rrrr}
    \toprule
    & & & & \multicolumn{4}{c}{\textbf{Hardware Feature Sensitivity}} \\
    \cmidrule(lr){5-8}
    \textbf{Param} & \textbf{Layer} & \textbf{Mode} & $\mathbf{\sigma}$ & \multicolumn{1}{c}{$T_1$} & \multicolumn{1}{c}{$T_2$} & \multicolumn{1}{c}{$\epsilon_g$} & \multicolumn{1}{c}{$\epsilon_r$} \\
    \midrule
    
    \multirow{4}{*}{\textbf{$\gamma$}} & \multirow{4}{*}{L1} 
      & M1 & 1.81 & -0.58 & \textbf{+0.81} & +0.09 & +0.08 \\
    & & M2 & 1.38 & +0.09 & -0.00 & -0.29 & \textbf{+0.95} \\
    & & M3 & 0.56 & \textbf{+0.80} & +0.55 & +0.23 & -0.01 \\
    & & M4 & 0.36 & -0.12 & -0.22 & \textbf{+0.93} & +0.29 \\
    \addlinespace

    \multirow{4}{*}{\textbf{$\gamma$}} & \multirow{4}{*}{L2} 
      & M1 & 0.42 & -0.49 & \textbf{+0.86} & -0.04 & +0.10 \\
    & & M2 & 0.25 & +0.30 & +0.06 & -0.16 & \textbf{+0.94} \\
    & & M3 & 0.15 & \textbf{+0.80} & +0.50 & +0.25 & -0.24 \\
    & & M4 & 0.07 & -0.18 & -0.08 & \textbf{+0.96} & +0.22 \\
    \addlinespace
    
    \multirow{4}{*}{\textbf{$\gamma$}} & \multirow{4}{*}{L3} 
      & M1 & 0.53 & -0.42 & \textbf{+0.76} & -0.14 & +0.47 \\
    & & M2 & 0.37 & +0.42 & -0.32 & -0.14 & \textbf{+0.84} \\
    & & M3 & 0.13 & \textbf{+0.78} & +0.53 & -0.24 & -0.23 \\
    & & M4 & 0.10 & +0.20 & +0.20 & \textbf{+0.95} & +0.14 \\
    \midrule

    \multirow{4}{*}{\textbf{$\beta$}} & \multirow{4}{*}{L1} 
      & M1 & 3.18 & -0.54 & \textbf{+0.83} & -0.07 & +0.13 \\
    & & M2 & 2.42 & +0.14 & -0.08 & -0.22 & \textbf{+0.96} \\
    & & M3 & 1.19 & \textbf{+0.83} & +0.56 & +0.03 & -0.07 \\
    & & M4 & 0.69 & -0.03 & +0.03 & \textbf{+0.97} & +0.23 \\
    \addlinespace

    \multirow{4}{*}{\textbf{$\beta$}} & \multirow{4}{*}{L2} 
      & M1 & 2.95 & -0.46 & \textbf{+0.77} & -0.09 & +0.44 \\
    & & M2 & 2.27 & +0.27 & -0.36 & -0.19 & \textbf{+0.87} \\
    & & M3 & 1.10 & \textbf{+0.84} & +0.53 & +0.07 & -0.02 \\
    & & M4 & 0.60 & -0.05 & -0.04 & \textbf{+0.98} & +0.21 \\
    \addlinespace

    \multirow{4}{*}{\textbf{$\beta$}} & \multirow{4}{*}{L3} 
      & M1 & 4.01 & -0.40 & \textbf{+0.74} & -0.11 & +0.53 \\
    & & M2 & 3.03 & +0.28 & -0.47 & -0.20 & \textbf{+0.82} \\
    & & M3 & 1.50 & \textbf{+0.88} & +0.48 & +0.01 & -0.02 \\
    & & M4 & 0.85 & +0.01 & -0.01 & \textbf{+0.97} & +0.23 \\
    \bottomrule
  \end{tabular}
\end{table*}

\begin{table*}[t]
  \centering
  \small
  \caption{\textbf{Z-Basis Dominant FiLM Modes ($d=11$).} The Feature columns show the sensitivity of the mode to specific hardware defects. Values represent the normalized singular vector component (signed), with dominant drivers highlighted in bold.}
  \label{tab:film_svd_z}
  
  \begin{tabular}{cc cc rrrr}
    \toprule
    & & & & \multicolumn{4}{c}{\textbf{Hardware Feature Sensitivity}} \\
    \cmidrule(lr){5-8}
    \textbf{Param} & \textbf{Layer} & \textbf{Mode} & $\mathbf{\sigma}$ & \multicolumn{1}{c}{$T_1$} & \multicolumn{1}{c}{$T_2$} & \multicolumn{1}{c}{$\epsilon_g$} & \multicolumn{1}{c}{$\epsilon_r$} \\
    \midrule
    
    \multirow{4}{*}{\textbf{$\gamma$}} & \multirow{4}{*}{L1} 
      & M1 & 0.43 & -0.07 & +0.31 & -0.03 & \textbf{+0.95} \\
    & & M2 & 0.19 & \textbf{+0.96} & -0.22 & -0.09 & +0.14 \\
    & & M3 & 0.12 & -0.20 & \textbf{-0.86} & +0.37 & +0.28 \\
    & & M4 & 0.05 & +0.18 & +0.34 & \textbf{+0.92} & -0.07 \\
    \addlinespace

    \multirow{4}{*}{\textbf{$\gamma$}} & \multirow{4}{*}{L2} 
      & M1 & 0.52 & -0.21 & -0.50 & +0.03 & \textbf{-0.84} \\
    & & M2 & 0.26 & \textbf{-0.95} & +0.08 & +0.25 & +0.20 \\
    & & M3 & 0.12 & -0.21 & -0.39 & \textbf{-0.86} & +0.25 \\
    & & M4 & 0.06 & -0.14 & \textbf{+0.77} & -0.44 & -0.44 \\
    \addlinespace

    \multirow{4}{*}{\textbf{$\gamma$}} & \multirow{4}{*}{L3} 
      & M1 & 0.65 & -0.41 & +0.41 & -0.06 & \textbf{+0.81} \\
    & & M2 & 0.22 & \textbf{+0.87} & -0.10 & -0.08 & +0.48 \\
    & & M3 & 0.06 & +0.28 & \textbf{+0.88} & +0.28 & -0.28 \\
    & & M4 & 0.03 & -0.04 & -0.24 & \textbf{+0.96} & +0.17 \\
    \midrule

    \multirow{4}{*}{\textbf{$\beta$}} & \multirow{4}{*}{L1} 
      & M1 & 2.59 & -0.13 & -0.25 & +0.09 & \textbf{-0.95} \\
    & & M2 & 1.50 & \textbf{+0.98} & -0.20 & -0.05 & -0.09 \\
    & & M3 & 0.57 & -0.15 & \textbf{-0.92} & +0.21 & +0.28 \\
    & & M4 & 0.29 & +0.10 & +0.22 & \textbf{+0.97} & +0.02 \\
    \addlinespace

    \multirow{4}{*}{\textbf{$\beta$}} & \multirow{4}{*}{L2} 
      & M1 & 2.87 & -0.04 & +0.28 & -0.07 & \textbf{+0.96} \\
    & & M2 & 1.68 & \textbf{+0.97} & -0.19 & -0.10 & +0.09 \\
    & & M3 & 0.70 & +0.20 & \textbf{+0.93} & -0.11 & -0.28 \\
    & & M4 & 0.33 & +0.12 & +0.11 & \textbf{+0.99} & +0.05 \\
    \addlinespace

    \multirow{4}{*}{\textbf{$\beta$}} & \multirow{4}{*}{L3} 
      & M1 & 3.45 & -0.04 & -0.25 & +0.08 & \textbf{-0.97} \\
    & & M2 & 2.09 & \textbf{+0.98} & -0.18 & -0.09 & +0.00 \\
    & & M3 & 0.84 & -0.16 & \textbf{-0.93} & +0.20 & +0.26 \\
    & & M4 & 0.39 & +0.12 & +0.20 & \textbf{+0.97} & +0.03 \\

    \bottomrule
  \end{tabular}
\end{table*}
\endgroup

\end{document}